# Electronic interactions in Dirac fluids visualized by nano-terahertz spacetime interference of electron-photon quasiparticles

Authors: Suheng Xu[1†], Yutao Li[1,2†], Rocco A. Vitalone[1], Ran Jing[1,3,4], Aaron. J. Sternbach[1,5], Shuai Zhang[1], Julian Ingham[1], Milan Delor[6], James. W. McIver[1], Matthew Yankowitz[7,8], Raquel Queiroz[1], Andrew J. Millis[1,9], Michael M. Fogler[10], Cory R. Dean[1], Abhay N. Pasupathy[1], James Hone[11], Mengkun Liu[3,12], D.N. Basov[1*]

Affiliations:

[1]*Department of Physics, Columbia University, New York, New York 10027, USA*

[2]*Brookhaven National Laboratory, Upton, New York 11973, USA*

[3]*Department of Physics and Astronomy, Stony Brook University, Stony Brook, New York 11794, USA*

[4]*Condensed Matter Physics and Materials Science Department, Brookhaven National Laboratory, Upton, New York 11973, USA*

[5]*Department of Physics, University of Maryland, College Park, Maryland 20742, USA*

[6]*Department of Chemistry, Columbia University, New York, New York 10027, USA*

[7]*Department of Physics, University of Washington, Seattle, Washington 98195, USA*

[8]*Department of Materials Science and Engineering, University of Washington, Seattle, Washington 98195, USA*

[9]*Center for Computational Quantum Physics, The Flatiron Institute, 162 5th Avenue, New York, New York 10010, USA*

[10]*Department of Physics, University of California at San Diego, La Jolla, CA 92093-0319, USA*

[11]*Department of Mechanical Engineering, Columbia University, New York, New York 10027, USA*

[12]*National Synchrotron Light Source II, Brookhaven National Laboratory, Upton, New York 11973, USA*

*Corresponding author: D. N. Basov: <u>db3056@columbia.edu</u>,

† co-first authors

Abstract

Ultraclean graphene at charge neutrality hosts a quantum critical Dirac fluid of interacting electrons and holes. Interactions profoundly affect the charge dynamics of graphene, which is encoded in the properties of its electronic-photon collective modes: surface plasmon polaritons (SPPs). Here we show that polaritonic interference patterns are particularly well suited to unveil the interactions in Dirac fluids by tracking polaritonic interference in time at temporal scales commensurate with the electronic scattering. Spacetime SPP interference patterns recorded in tera-hertz (THz) frequency range provided unobstructed readouts of the group velocity and lifetime of polariton that can be directly mapped onto the electronic spectral weight and the relaxation rate. Our data uncovered prominent departures of the electron dynamics from the predictions of the conventional Fermi-liquid theory. The deviations are particularly strong when the densities of electrons and holes are approximately equal. The proposed spacetime imaging methodology can be broadly applied to probe the electrodynamics of quantum materials.

## Main text

Plasmon polaritons are coherent electron density oscillations dressed with electromagnetic fields. These collective modes play a prominent role in the charge response of metals and semiconductors. Graphene is an exemplary platform for plasmonics due to its high tunability and relatively low electronic losses(*1–4*). In most previous studies of plasmon polaritons in graphene, the graphene was heavily doped, so that many-body effects did not qualitatively alter the conventional Fermi-liquid picture of charge dynamics(*3*). However, electrons in graphene near the charge neutral point (CNP) form a correlated Dirac fluid(*5–7*), which exhibits unusual behaviors such as a quantum critical scattering rate, breakdown of Wiedemann-Franz law, and large magnetoresistance(*8–12*). These exotic properties stem from electron-electron interactions, whose strength is parametrized by the ratio between Coulomb energy and kinetic potential: the effective fine-structure constant $\alpha = e^2/\epsilon \hbar v_F$ (*13–15*). Here, $\epsilon$ is the dielectric constant of the environment and $v_F$ is the graphene Fermi velocity.

To investigate the collective behavior of the Dirac fluid, we visualized the "worldlines" of surface plasmon polaritons (SPPs), i.e., the trajectories of polaritonic wave packets in spacetime. The experiments were performed using a home-built cryogenic terahertz scanning near-field optical microscope (THz-SNOM)(*16–18*) that gave us access to sub-picosecond dynamics with sub-50-nm spatial resolution. From the measured SPP worldlines we extracted two key observables, the SPP group velocity $v_g$ and the SPP lifetime $\tau_p$, that provided evidence for non-Fermi-liquid effects in the charge dynamics.

## Nano-THz response of a graphene ribbon

We investigated graphene ribbons encapsulated in hexagonal boron nitride (hBN). The ribbons were integrated in a back-gated structure assembled on a Si/SiO$_2$ wafer. Ohmic Cr/Au contacts enabled electrical measurements (Fig. 1a, Supplemental material S1). Our THz-SNOM operated with broadband THz pulses, spanning a range of frequencies from 0.5-1.5 THz (Fig. 1a), generated via a photo-conductive antenna (PCA). The THz pulses were focused on the tip of an atomic force microscope (AFM) and the forward-scattered THz pulses were detected by another PCA with sub-picosecond (ps) temporal resolution. The demodulation of the detected signal produced artifact-free THz images with a 50-nm spatial resolution as described in Refs. (*16*, *17*)(Supplemental material S2). By raster scanning the sample under the tip, we obtained images of the amplitude and phase of the local THz electric field(*18*).

A representative image of a nano-THz signal is shown in Fig. 1b. The false color plot displays the amplitude of the local field. The nano-THz signal on graphene is enhanced compared to that of the substrate owing to the higher THz conductivity. The nano-THz response is maximized at the center of the graphene ribbon. This spatially inhomogeneous response originates from the graphene SPP, as we detail below. To help us analyze the data, we calculated the THz near-field reflection coefficient ($r_p$) for our structure, which is composed of encapsulated graphene, 285-nm-thick $SiO_2$ and the Si back gate. (Fig.1c). The SPP dispersion manifests itself as a line of enhanced Im $r_p$(*19*) in the frequency-momentum parameter space (the red dashed line in Fig. 1c). The nearly linear (rather than common square-root-like) dispersion of the graphene SPP in the parameter space accessible with THz nano-imaging is caused by the screening from the back gate. (Supplemental material S3).

To complete the basic nano-THz characterization of our structure, we investigated the evolution of the response of graphene as we tuned its Fermi level with the gate voltage $V_{bg}$. In Fig. 1d,e, we show data obtained by scanning across the graphene ribbon (along the dashed line in Fig. 1b) while varying $V_{bg}$. The amplitude of the nano-THz signal increases with the gate voltage as the Fermi level is driven away from the CNP and the graphene THz conductivity increases(*20*). The CNP of the sample is at $V_{bg}$~ 0 V, attested by the symmetric gate dependence in Fig. 1d,e and further confirmed by transport measurements (Supplemental material S4). At low carrier densities ($|V_{bg}| < 3$ V), both the amplitude and phase of the nano-THz signal are maximized at the center of the ribbon. However, at $|V_{bg}| > 3$ V, the maximum of the phase signal shifts to the edges, whereas the amplitude signal remains most prominent at the center. These observations can be explained by the change in the SPP wavelength(*18*) (Supplemental material S8): namely, our sample acts as a plasmonic cavity supporting multiple reflections at the ribbon boundaries, which create wavelength-dependent interference patterns in the interior of the cavity. As we sweep the back gate voltage from $|V_{bg}|$=0 V to $|V_{bg}|$=30 V, the calculated wavelength of THz SPPs grows from 5 to 26 μm, the latter number exceeding the width of the ribbon.

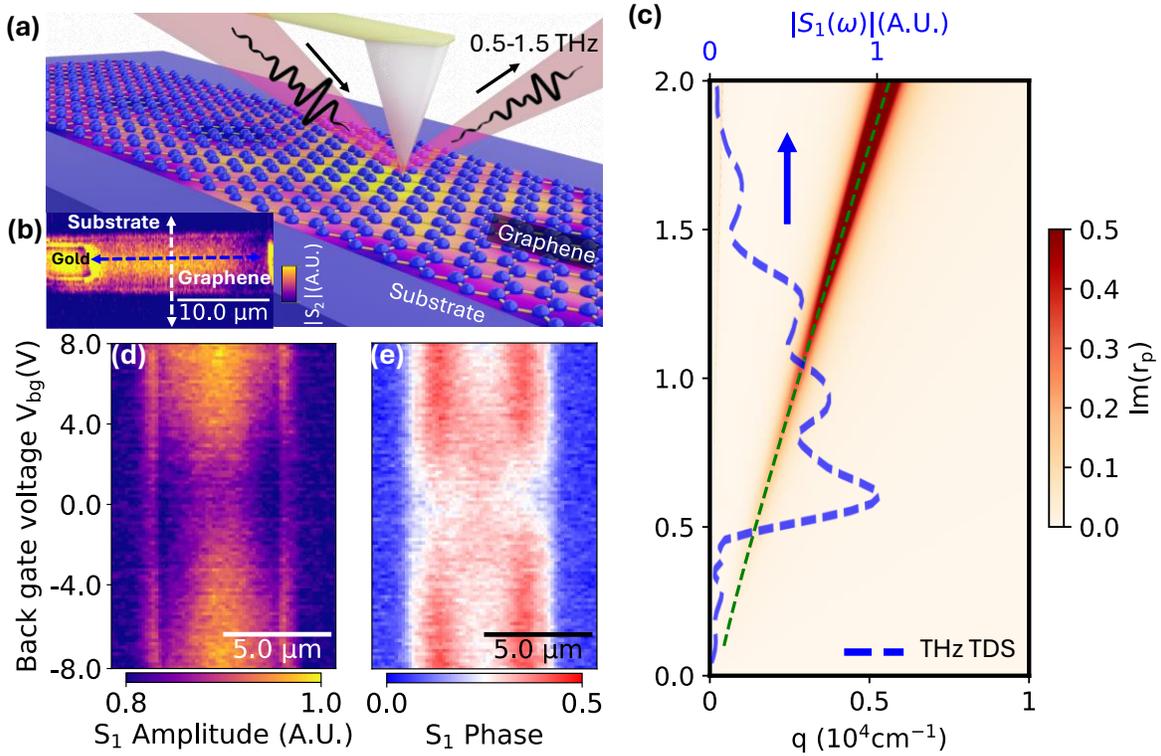

**Fig. 1. Nano-THz imaging of plasmons in a gateable graphene cavity.** (a) Schematics of the nano-THz experiment setup. A monolayer graphene ribbon encapsulated by hBN resides on a SiO$_2$/Si substrate and is illuminated with broadband terahertz pulses (0.5-1.5 THz). The tip acts as a nano-antenna, launching and detecting polaritonic wave packets. (b) Nano-THz image of the broadband THz signal on a graphene ribbon at $V_{bg}$=5 V and T=297 K. (c) The imaginary part of the reflection coefficient ($r_p$) calculated as a function of momentum and frequency at carrier density $n = 1.1 \times 10^{12}\ cm^{-2}$. The maximum of Im($r_p$), indicated by the green dashed line, traces the SPP dispersion, which is nearly linear in momentum. The spectrum of the nano-THz pulse is plotted as a blue dashed line. (d,e) Gate voltage dependence of nano-THz amplitude and phase profile across the graphene ribbon, obtained by scanning along the white dashed line marked in (b).

## Spacetime mapping of electron-photon polaritons

Tracking the moving polaritonic wave packets necessitates high spatial and temporal resolution. For example, the spatiotemporal dynamics of infrared hyperbolic phonon polariton in hBN was visualized by ultrafast transmission electron microscopes and scattering-type SNOMs(s-SNOM) integrated with time-domain interferometry(*21*, *22*). Terahertz waves traveling in graphene metamaterials have also been visualized using aperture-based THz near-field probe and spatially resolved electro-optical sampling, achieving a spatial resolution of a few micrometers(*23*, *24*). Inspired by these innovations, we devised and implemented nano-THz spacetime imaging offering a direct access to the electric field profiles of polariton wave packets with the spatial resolution down to $10^{-4}$ of the THz wavelengths in free space.

Here, we employed our THz-SNOM, which incorporates the terahertz time-domain spectroscopy(thz-tds) and s-SNOM, simultaneously offers exceptional spatial and temporal resolution to record the SPP dynamics in graphene on a map with ultrafine spacetime pixel. A representative map collected at $V_{bg} = 15\ V$ and $T = 297$ K is presented in Fig. 2b. To obtain this map, we scanned the tip along the white dashed line in Fig. 1b while slowly varying the time delay, thus unfolding the temporal profile of the scattered THz pulses (Supplemental material S5). The signal along the vertical time axis manifests itself as a time domain trace of the tip-scattered THz pulse. One typical trace shown in Fig. 2a is extracted from the center of the graphene ribbon (dashed black line in Fig. 2b). To eliminate the spatially uniform background, we applied the second spatial derivative to this spacetime map. As shown in Fig. 2c (also Supplemental material S6), the derivative plot reveals a periodic checkerboard pattern.

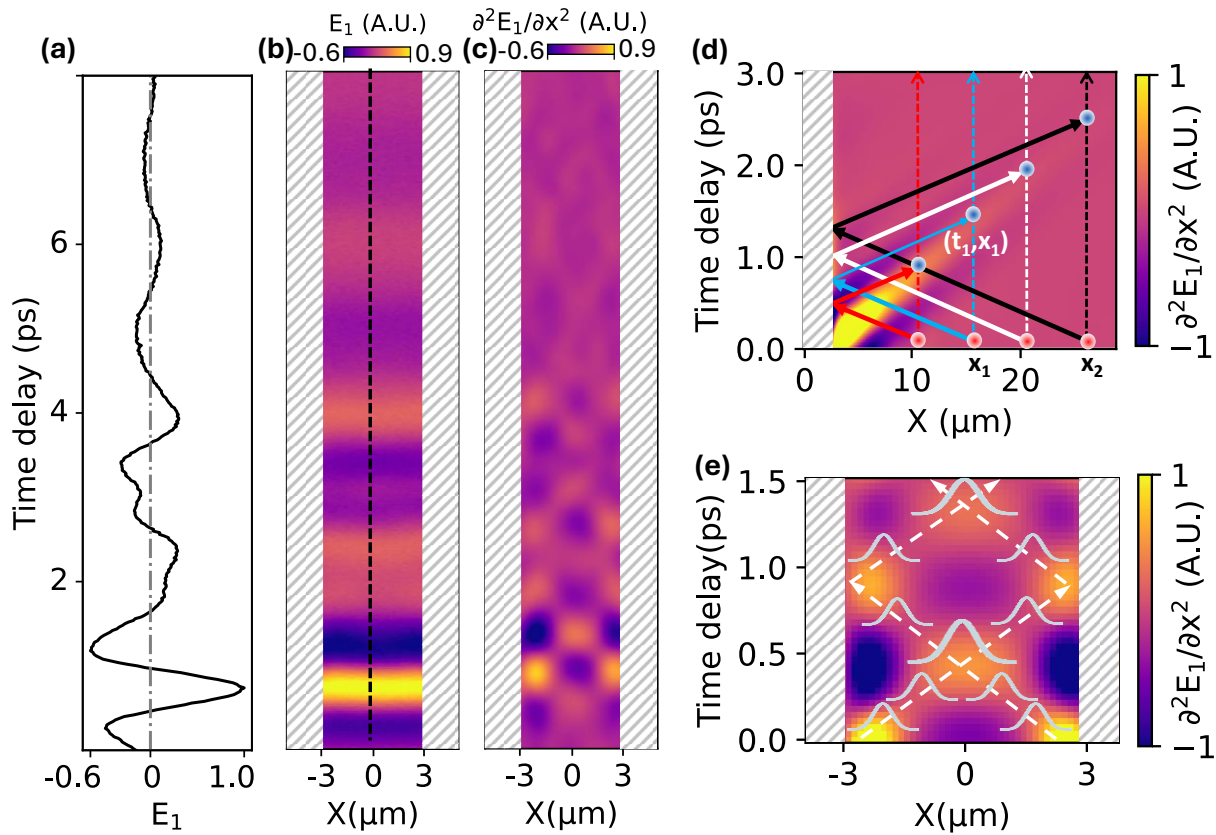

**Fig. 2. Spacetime mapping of plasmon polaritons in a graphene ribbon cavity.** (a) THz time domain signal collected with the tip located at the center of the graphene ribbon. (b) Nano-THz spacetime map. The data are obtained by scanning the tip along the same line across the graphene sample while sweeping the time delay between the scattered THz pulse and the optical detection beam. The substrate region is indicated by the gray shaded area. (c) Second spatial derivative of the data in panel (b), showing the checkerboard pattern. This pattern is generated by SPP wave packets launched by the tip as they propagate across the ribbon and reflect off its edges. (d) Simulated spacetime map in a half-plane sample where SPPs are launched at a time delay of 0 ps by the tip, located at various distances X from the edge. The launching events are labeled with the red dots and the detection events with the blue dots. The solid arrows whose slope is equal to the reciprocal of the SPP group velocity show the plasmon trajectories. The coalescence of detection event traces forms the worldline of the SPPs. In these numerical simulations, the group velocity is 20 µm/ps and the plasmon lifetime is 0.85 ps. (e) Simulated nano-THz spacetime map of SPPs in a ribbon cavity. The substrate region is marked by the shaded area. The dashed white lines mark the SPP detection worldline.

The formation of the checkerboard pattern can be understood from the geometrical construction shown in Fig.2d. Consider an SPP launching event at t=0 ps by the tip located at a distance $x_1$ from the edge. Since the SPP dispersion is almost linear (Fig. 1e), various frequency components of the broadband THz pulse create polaritons propagating with the same speed $v_g$. The launched SPP wave packet travels outward from the tip, undergoes a reflection from the edge, and then travels inward. If the propagation losses are not too large, so that the reflection wave packet is able to complete the round trip, this event is detected by the THz-SNOM at time $t_1 = 2x_1/v_g$. In Fig. 2d, we depict a few such launching and detection events (red and blue dots) as well as the corresponding round-trip SPP trajectories (solid lines with slopes $\pm v_g^{-1}$). As the tip is moved away from the edge, the detection time grows proportionally longer, producing an SPP "detection worldline" in the spacetime coordinates. This worldline emanates from the point $(x = 0, t = 0)$ and has slope $2/v_g$. The worldline fades away at large distances and long times, providing information about plasmonic losses. A more comprehensive elucidation of the spacetime map is provided in Supplemental material S5.

In the above discussion, we considered the SPP reflections only from the edge nearest to the tip. In Fig. 2e, we model the spacetime map for the case of a finite-width plasmonic cavity where both edges can reflect the SPPs, so that each of them generates a worldline. The worldlines begin at $t = 0$ at the two opposite edges and tilt inward to intersect at the center. At the intersection point, the SPP wave packets reflected from the two edges return to the tip simultaneously. The worldlines reverse their slope once they reach an edge of the ribbon; this corresponds to repeated reflections of the wave packets inside the cavity. Due to the finite width of the wave packets, the net observable pattern resembles a checkerboard pattern, in agreement with the experimental data in Fig. 2c.

## SPP dynamics and worldlines

To study the dynamics of the SPP wave packets, we obtained spacetime maps at various temperatures at a fixed back gate voltage. In Fig.3a-c, we present three spacetime maps collected at $T = 300$ K, 170 K, and 55 K (all at $V_{bg} = 15$ V). In the spacetime maps, the SPPs at lower temperatures show more robust and longer-lived oscillations. The temporal oscillations of the SPPs extracted from the center of the spacetime map (white dashed lines in Fig. 3a-c) are summarized in Fig. 3d. These oscillations can be modeled as described in Supplemental material S9. The best fit to the experimental data is found when the simulated plasmon lifetime is 0.75 ps, 1.8 ps, and 4.75 ps (Fig. 3d). These plasmon lifetimes correspond to plasmonic relaxation rates ($1/\tau_p$) of 1.33, 0.56, and 0.21 THz, respectively. Except for the first one, these lifetimes are longer than the previously measured record-high lifetime of infrared SPPs(*3*) (1.6 ps at $T = 60$ K and photon energy $\sim$110 meV(*3*)).

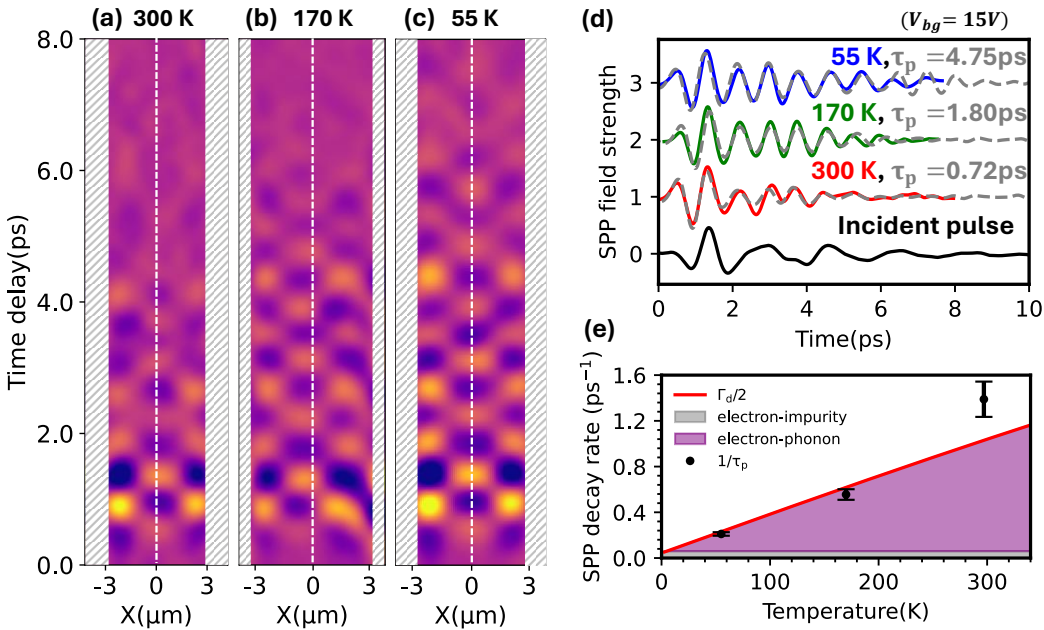

**Fig. 3. Temperature dependence of the SPP relaxation rate in the Fermi liquid regime.** (a-c) Second-order spatial derivatives of spacetime maps taken at 300 K, 170 K and 55 K, all at gate voltage $V_g =15$ V. (d) Temporal oscillations of the SPP field at the center of the graphene cavity along the white dashed lines in a-c. The gray dashed lines are simulated temporal oscillations yielding the SPP lifetimes of 0.75 ps (T=300 K), 1.8 ps (T=170 K) and 4.75 ps (T=55 K). The black solid line is the THz scattering signal probed at the location of the gold launcher. (e) Temperature dependence of the SPP decay rate (inverse lifetime). The black dots are the SPP decay rates $1/\tau_p$

extracted from the fittings in panel (d). The red curve is the estimate of the SPP relaxation rate, considering various momentum-relaxing scattering mechanisms as described in the text. The shaded grey region shows the relaxation rate contributed by electron-impurity scattering, whereas the magenta region depicts the contribution of electron-phonon scattering. The scattering rate analysis is described in Supplemental material S11.

The spacetime patterns of polaritonic wave packets (Figs.2-3) are ultimately governed by the dispersion of SPP in the frequency ($\omega$)-momentum (q) space. The SPP dispersion can be derived from the pole of the near-field reflection coefficient via $r_p(q,\omega) \simeq 1 - (2qd/\epsilon_{2D})$ for q $\gg \omega/c$. Here, $\epsilon_{2D} = 1 + (iq^2/\omega)e^2 V\sigma$ is the 2D dielectric function of graphene, $V = V(q,\omega)$ is the effective Coulomb interaction kernel, and $\sigma$ is the sheet conductivity(25). Under our experimental conditions, $V$ is well approximated by the inverse capacitance $V = C^{-1} = 4\pi d/\epsilon$ per unit area, where $d$ is the distance between the graphene and the gate and $\epsilon$ is the dielectric constant of the spacer. The nontrivial physics we observe is encoded in the optical conductivity $\sigma = \sigma(\omega)$, accessible through the analysis of SPP worldlines and dispersions.

In the Fermi liquid regime realized in heavily doped graphene with the chemical potential μ exceeding the temperature $\mu \gg k_b T$, the conductivity is well captured by the Drude formula:

$$\sigma(\omega) = \frac{i}{\pi}\frac{D}{\omega + i\Gamma_d}, \qquad (1)$$

where $D$ is the Drude weight and $\Gamma_d$ is the momentum-relaxing scattering rate(26). Assuming $\omega \gg \Gamma_d$, the corresponding observables in spacetime maps read as:

$$v_g = \sqrt{\frac{D}{\pi C}}, \qquad \tau_p = \frac{2}{\Gamma_d} \qquad (2)$$

The SPP lifetime ($\tau_p$) is twice as large as the invers of $\Gamma_d$, which can be attributed to the screening effect produced by the back gate. (see Supplemental section S10 for details). Hence, the two main characteristics of the SPP worldlines -- the slope ($2/v_g$) and the decay time ($\tau_p$) -- are governed in the Drude model by the Drude weight and the scattering rate, respectively. Both quantities can be directly measured as the geometrical metrics of worldlines. The calculated SPP decay rate (red curve in Fig.3e) shows that a combination of electron-phonon and impurity scattering is in quantitative agreement with the SPP-extracted data, as detailed in Supplemental material S11.

**Electronic interactions from spacetime metrology**

We now focus on the SPP spacetime dynamics in the vicinity of the CNP. Our THz spacetime maps collected at ambient temperature reveal plasmonic propagation at gate voltages as low as 5 V, 1 V or even 0 V at the CNP (Fig.4). We attribute the presence of SPPs at the CNP to the response of thermally excited electrons and holes(*9, 10, 27, 28*). The SPP group velocity increases with gate voltage, as evidenced by shallower worldline slopes (dashed arrows in Fig. 4a-c and Fig. 4e-g). The evolution of the group velocity as a function of the back gate voltage is summarized in Fig. 4d. The group velocity of SPP increases with carrier density from 6 μm/ps or 2% of the speed of light near the CNP to 26 um/ps at $V_{bg} = 30$ V. The worldline of the SPPs at the CNP is short-lived: the plasmonic wave packets cease to exist after only 0.2-0.3 ps. Applying a gate voltage as low as 1 V nearly doubles the lifetime manifested by the extended SPP propagation range. We quantify the SPP lifetime by analyzing the temporal decay of the SPP field along the worldlines of SPP that travel along the ribbon at various gate voltages (Fig. 4h, with additional details presented in Supplemental material S14). The SPP relaxation rate decreases from ~5 THz at the CNP to less than 1.5 THz at $\mu > 120$ meV (Fig. 4i). The significant enhancement of the SPP relaxation rate at the CNP cannot be explained solely by the momentum-relaxing scattering rates, which generally either remain unchanged or decrease with carrier density(*29*). The influence of the interband transition in nano-THz data is also negligible (Supplemental material S13).

Searching for the origins of the evolution of the SPP dynamics in Fig. 4d,i, we recall that in the absence of Umklapp processes, dc charge transport in monopolar Fermi liquids is not impeded by momentum-conserving electron scattering because the total momentum and current are proportional to each other. This is the case in highly doped graphene. However, the situation is different in materials hosting both electrons and holes(*14, 30–32*) such as charge-neutral graphene at a finite temperature. The electron-hole scattering produces current relaxation and therefore enhanced plasmonic losses(*30, 33–35*). Below, we attempt to verify the importance of this so-called electron-hole drag effect by examining how the conductivity we deduce from our measurements varies as a function of the gate voltage at a finite temperature.

For the quantitative analysis of the doping dependence trends in Fig.4, we have adopted the two-component model of the THz conductivity of graphene(*36*):

$$\sigma_h(\omega) = \frac{i}{\pi} \frac{D_h}{\omega + i\Gamma_d} + \frac{i}{\pi} \frac{D_k - D_h}{\omega + i\Gamma_d + i\Gamma_e}, \qquad (3)$$

where $D_h$ is the so-called hydrodynamic Drude weight and $D_k$ is the kinetic Drude weight; $\Gamma_d$ is the momentum-relaxing scattering rate attributed to the combined action of impurities and phonons; and $\Gamma_e$ is the electronic scattering rate (Supplemental material S14). The first term in (3) describes the contribution from the hydrodynamic-type flow, where all the charge carriers move in the same direction. Accordingly, $D_h$ is nonzero only away from the CNP. The second term in (3) represents the contribution associated with the electrons and holes moving in opposite directions in response to the same electric field. The electron-hole drag is manifest in the added momentum relaxation rate $\Gamma_e$ of this term. We display the ratio of the spectral weights (i.e., the numerators) of the two contributions in the inset of Fig. 4d. This plot shows that the hydrodynamic Dirac fluid behavior dominates the graphene response at elevated temperatures and at lower carrier densities.

As shown in Fig. 4d,i, the two-component model [Eq. (3)] captures the trends seen in the evolution of both the scattering rate and the group velocity as functions of the chemical potential. For simplicity, we assumed that the scattering rates $\Gamma_d$ and $\Gamma_e$ are constant and obtained the best fit to the data for $\Gamma_e = 3.0$ THz and $\Gamma_d = 0.47$ THz. These choices of parameters reproduce the entire gate-dependent data set developed from 17 separate spacetime maps. We remark that in a more accurate model, $\Gamma_e$ should decrease away from the CNP as the system evolves into a liquid with a progressively larger Fermi surface(*37*). At the CNP, the hydrodynamic term in Eq. (3) vanishes, so that the current relaxation rate reaches its maximum value of $\Gamma_d + \Gamma_e$. At the CNP, the electronic scattering time $\Gamma_e^{-1}$ is actually shorter than a single oscillation cycle of the SPPs, yet the effect is still observable in our data. Simultaneously, the SPP group velocity $v_g$ attains its minimum at the CNP (blue and red lines in Fig.4d). Evidently, electronic interactions slow down the SPP in a Dirac fluid in a manner analogous to a viscous drag. From the fitted $\Gamma_e$, we estimate the graphene fine-structure constant as 0.34 (Supplemental material S17), which matches the theoretical estimate (assuming $v_F = 1.3 \times 10^6$ m/s and $\kappa = 5$).

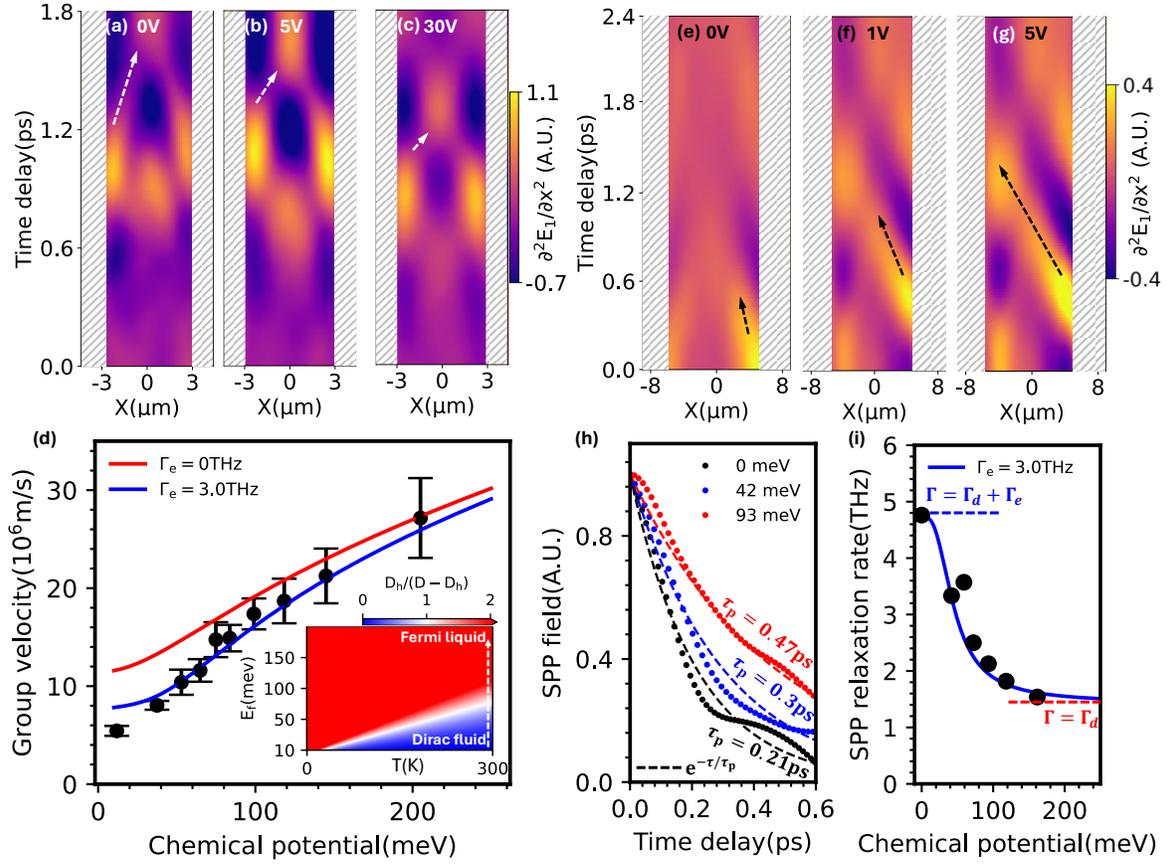

**Fig. 4. SPP dynamics renormalized by electronic interactions in Dirac fluid** (a-c) Spacetime maps collected across the graphene ribbon when applying different back gate voltages. White arrows mark the slopes of the SPP worldline. A shallower slope corresponds to a faster SPP group velocity. (d) Dots display the gate dependence of the graphene SPP group velocity at 300K. Solid red line: theoretical group velocity obtained from Eq. (3) by setting $\Gamma_e = 0$ THz, as described in the text. The solid blue trace was obtained by setting the electronic scattering rate $\Gamma_e = 3$ THz. The false color map in the inset of (d) presents the ratio between the two numerators $D_h$ and $D_k - D_h$ in the two-component model [Eq. (3)]. The white dashed line presents the trajectory of gate-dependent measurements probing the transition from a Dirac fluid regime to a Fermi liquid regime. (e-g) Spacetime maps collected along the graphene ribbon (marked as blue dashed lines in Fig. 1b) at $V_g$=0 V (e), 1 V( f), and 5 V (g). The black arrows mark the SPP worldlines. (h) Time-dependent polariton field strength extracted along the SPP worldlines in (e-g). The field strength is normalized by its maximum value. The dashed curves show the exponential decay $e^{-t/\tau_p}$ controlled by the SPP lifetime $\tau_p$. (i) Dirac-fluid/Fermi-liquid crossover visualized by the gate-dependent SPP relaxation at room temperature. In the Dirac fluid regime ($E_f \ll k_b T$), the SPP damping is determined by $\Gamma_d + \Gamma_e$ (marked by the blue dashed line). In the Fermi liquid regime ($E_f \gg k_b T$), the SPP damping is solely determined by $\Gamma_d$ (marked by the red dashed line). The blue solid line shows the calculated gate-dependent SPP relaxation rate based on the two-fluid model ($\Gamma_d = 0.47$ THz, $\Gamma_e = 3$ THz).

To summarize, we introduced a nano-THz spacetime metrology to measure the group velocity and lifetime of SPP wave packets in graphene. The developed spacetime mapping method grants access to the low-energy electrodynamics of quantum materials with sub-diffraction and sub-cycle spacetime resolution on the order of $50\text{nm} \times 50\text{fs}$. Our observations revealed a significant renormalization of both parameters near the charge neutrality point of graphene, which is consistent with the electron-hole drag effect in the Dirac fluid. Our nano-THz spacetime mapping opens a new realm for spatiotemporal control of polaritons(*21*, *22*, *24*, *38–46*) and for nano-spectroscopy of low-energy collective modes in many other quantum materials(*47–53*).

# Acknowledgements

**Funding**: The development of space-time metrology is supported as part of Programmable Quantum Materials, an Energy Frontier Research Center funded by the U.S. Department of Energy (DOE), Office of Science, Basic Energy Sciences (BES), under award DE-SC0019443". Research on electronic interactions in graphene is supported by DOE-BES DE-SC0018426. DNB is the Moore investigator in quantum Materials EPIQS GBMF9455 and Vannevar Bush Faculty Fellow ONR-VB: N00014-19-1-2630. S.X., R.A.V, R.J., M.K.L and D.N.B. acknowledge support for THz-SNOM development from the US Department of Energy (DOE), Office of Science, National Quantum Information Science Research Centers, Co-design Center for Quantum Advantage (Contract No. DE-SC0012704). M.K.L. acknowledges support from the NSF Faculty Early Career Development Program under Grant No. DMR - 2045425.

**Author contributions**: D. N. B. conceived of the study. S. X. recorded the near-field data with assistance from R. A. V., R. J., A. J. S. and S. Z.; Y. L. prepared the samples, with guidance from J. H and C. R. D; S. X performed theoretical calculations and numerical simulations, with assistance from R. J., J. I., R. Q., A. J. M., M. M. F.; S. X. analyzed the data with assistance from  R. A. V., R. J., A. J. S, S. Z., M. D., J. W. M, M. Y., A. J. M, M. M. F, M. L.; S. X and D. N. B wrote the manuscript with input from all the coauthors.

**Competing interests**: The authors declare that they have no competing interests.

**Data and materials availability**: All the data that support the findings of this study are available from the corresponding authors upon reasonable request.

Reference

Supplemental material for

# Electronic interactions in Dirac fluids visualized by nano-terahertz spacetime interference of electron-photon quasiparticles

Authors: Suheng Xu[1][†], Yutao Li[1,2][†], Rocco A. Vitalone[1], Ran Jing[1,3,4], Aaron. J. Sternbach[1,5], Shuai Zhang[1], Julian Ingham[1], Milan Delor[6], James. W. McIver[1], Matthew Yankowitz[7,8], Raquel Queiroz[1], Andrew J. Millis[1,9], Michael M. Fogler[10], Cory R. Dean[1], Abhay N. Pasupathy[1], James Hone[11], Mengkun Liu[3,12], D.N. Basov[1*]

Affiliations:

[1]Department of Physics, Columbia University, New York, New York 10027, USA

[2]Brookhaven National Laboratory, Upton, New York 11973, USA

[3]Department of Physics and Astronomy, Stony Brook University, Stony Brook, New York 11794, USA

[4]Condensed Matter Physics and Materials Science Department, Brookhaven National Lab, Upton, New York 11973, USA

[5]Department of Physics, University of Maryland, College Park, Maryland 20742, USA

[6]Department of Chemistry, Columbia University, New York, New York 10027, USA

[7]Department of Physics, University of Washington, Seattle, Washington 98195, USA

[8]Department of Materials Science and Engineering, University of Washington, Seattle, Washington 98195, USA

[9]Center for Computational Quantum Physics, The Flatiron Institute, 162 5th Avenue, New York, New York 10010, USA

[10]NDepartment of Physics, University of California at San Diego, La Jolla, CA 92093-0319, USA

[11]Department of Mechanical Engineering, Columbia University, New York, New York 10027, USA

[12]National Synchrotron Light Source II, Brookhaven National Laboratory, Upton, New York 11973, USA

*Corresponding authors: D. N. Basov: db3056@columbia.edu,

[†]co-first authors

Table of contents:

# S1: Sample fabrication and geometry

Encapsulating graphene with two hexagonal boron nitride (hBN) layers is known to significantly improve the carrier mobility of graphene devices(*54*). To fabricate such a device, hBN and graphene flakes are first mechanically exfoliated onto $SiO_2$/Si wafer chips. Then, using a polypropylene carbonate (PPC) transfer slide, an hBN-graphene-hBN heterostructure is assembled. The top hBN needs to be thin (<10nm) so that the s-SNOM can effectively probe the electromagnetic response of the graphene underneath. However, thin hBN flakes are difficult to pick up using PPC. To get around this, the topmost hBN is a thick (~30nm) flake with a thin (~7nm) appendage. The presence of the thick part of the flake allows for relatively easy pickup by the PPC transfer slide. The graphene and bottom hBN flakes overlap with both the thick and thin parts of the top hBN, although only the area underneath the thin part of hBN is used in the actual device.

The hBN-graphene-hBN stack, residing on the PPC transfer slide, is released onto a Piranha-cleaned $SiO_2$/doped Si chip with alignment markers at 115°C. During the release process, special care is taken to remove the "bubbles" trapped between the layers of the stack by repeatedly expanding/shrinking the frontier of PPC/stack contact at 80°C. To remove the remaining PPC film, the $SiO_2$/Si chip is annealed at 360°C, $<10^{-5}$ mbar for 30 minutes. E-beam lithography is used to define the ribbon shape of the device, with bilayer 495 A4/950 A2 polymethyl methacrylate (PMMA) as the etch mask. Reactive ion etching (RIE) using $CHF_3$/Ar and $O_2$ removes the heterostructure outside the etch mask. After etching, the remaining PPC is dissolved by acetone, leaving the ribbon behind. Finally, e-beam lithography and e-beam metal evaporation are used to fabricate a pair of edge contacts to graphene that lead to two 200 $\mu m \times$ 200 $\mu m$ metal pads. The two pads and the doped Si are connected to pins on the sample holder by gold wires glued with silver paste on both ends.

During the measurement, the graphene is grounded while the doped Si is supplied with a DC bias $V_{bg}$. Based on the parallel capacitor model, the average carrier density in graphene can be determined from $V_{bg}$ and the thickness and permittivity of the dielectric spacer .

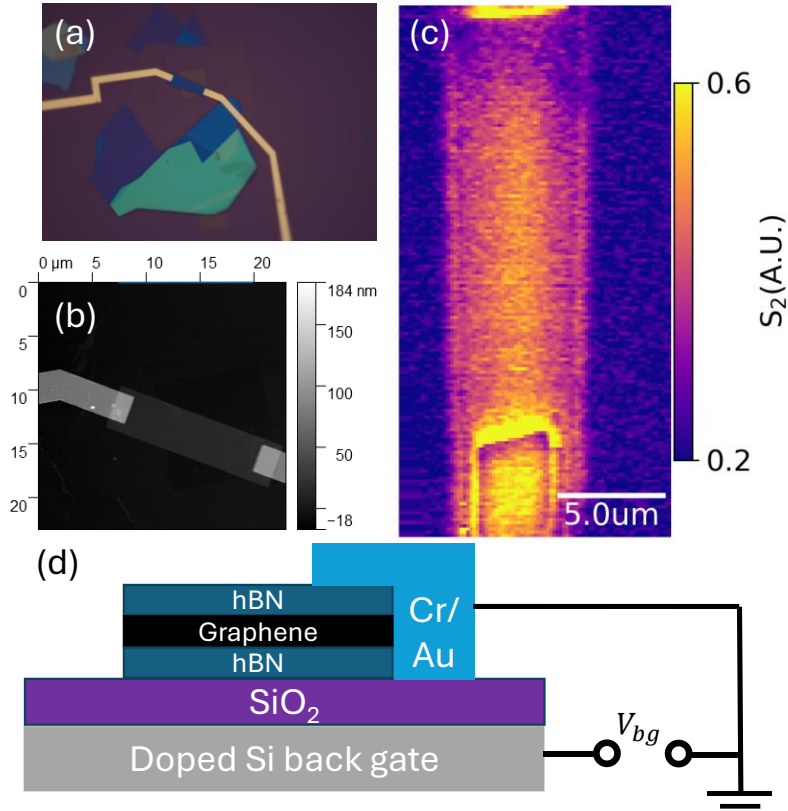

Fig. S1 (a) Optical image of the sample. (b) Atomic force microscopy measurement on the sample. (c) WL signal of the nano-THz measurement on the sample. (d) Schematic cartoon of the device.

## S2. THz scanning near-field optical microscopy

We conducted all the terahertz near-field measurements utilizing the home-built cryogenic THz scanning near-field optical microscope(*16–18*). The THz broadband pulse is generated and detected by a pair of photo-conductive antennas (PCAs, Menlo Systems GmbH). Our home-built cryogenic ultra-high vacuum atomic force microscope is operated in the tapping mode, in which the cantilever of a metallic tip (Rocky Mountain Nanotechnology, LLC) vibrates near its fundamental resonance frequency (30-80kHz). The THz radiation from the PCA emitter is collimated by a TPX lens and focused onto the tip and sample by a parabolic mirror. The scattered field is detected by an unbiased PCA and the photocurrent signal is demodulated by a lock-in amplifier at harmonics of the tip tapping frequency. Through this demodulation, we effectively probe the near-field tip-sample interaction with a spatial resolution on the order of 50nm.

## S3 Screened and free-standing THz plasmons in graphene

In the Terahertz frequency range, free carriers in a doped Si substrate screen the plasmon polaritons in graphene(18). Thus, it is necessary to include the doped substrate in the dispersion calculation of surface plasmon polaritons. In Fig. S2, we show the imaginary parts of p-polarized reflectance that use a doped (a) and an undoped (b) Si substrate. The carrier density in the doped Si substrate is $5 \times 10^{22} cm^{-3}$. In Fig. S2(a), the dispersion of surface plasmon polaritons is linear because of the screening. When the plasmon is unscreened, we see a familiar 2d plasmon dispersion, $\omega \sim \sqrt{q}$, as shown in Fig. S2(b).

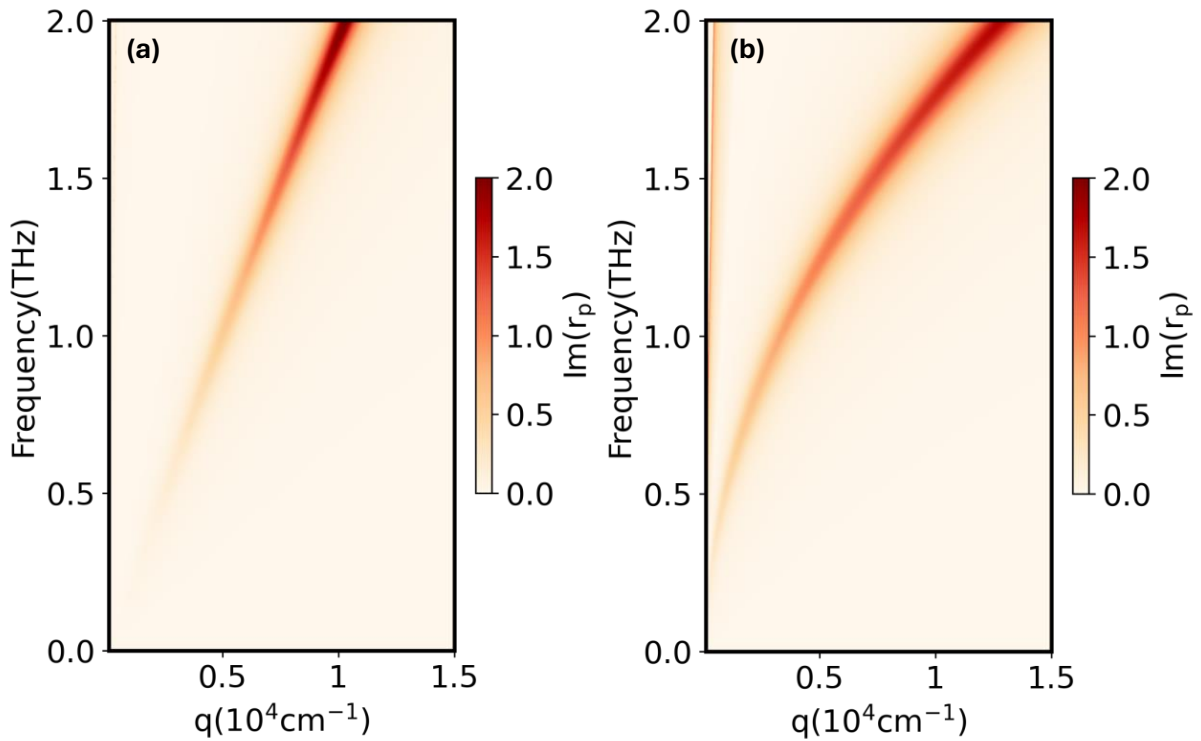

Fig. S2 Imaginary part of the p-polarized reflectance coefficient modeled for (a) screened and (b) free-standing graphene with chemical potential 0.1ev. In (a), we consider screening by a doped Si substrate with $n = 5 \times 10^{22} cm^{-3}$. In (b), we consider an undoped intrinsic Si substrate. A typical 2d plasmon dispersion, $\omega \sim \sqrt{q}$, shows up in the unscreened case. The sample structure for the above rp calculation is graphene/SiO2(300nm)/Si.

## S4. Gate-dependent resistivity

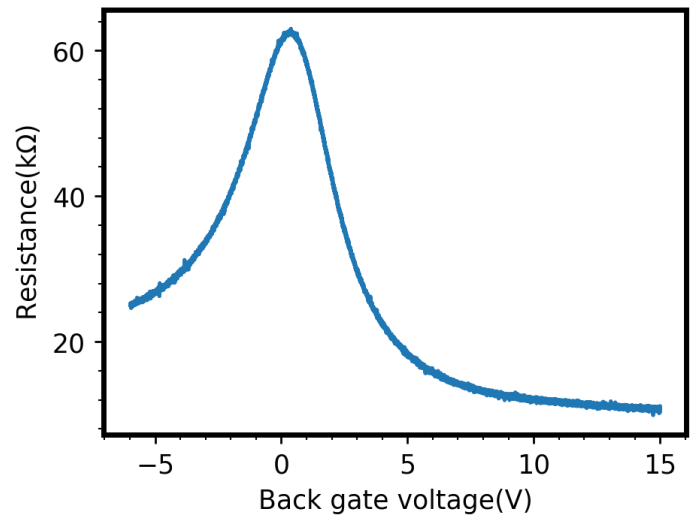

Fig. S3 In-situ measurement of graphene resistivity at room temperature. The resistance is determined by measuring the values at the two gold contacts located at the ends of the graphene ribbon. A peak in resistance serves as a distinctive indicator of reaching the charge neutrality point.

## S5 Spacetime maps and propagating surface plasmon polaritons

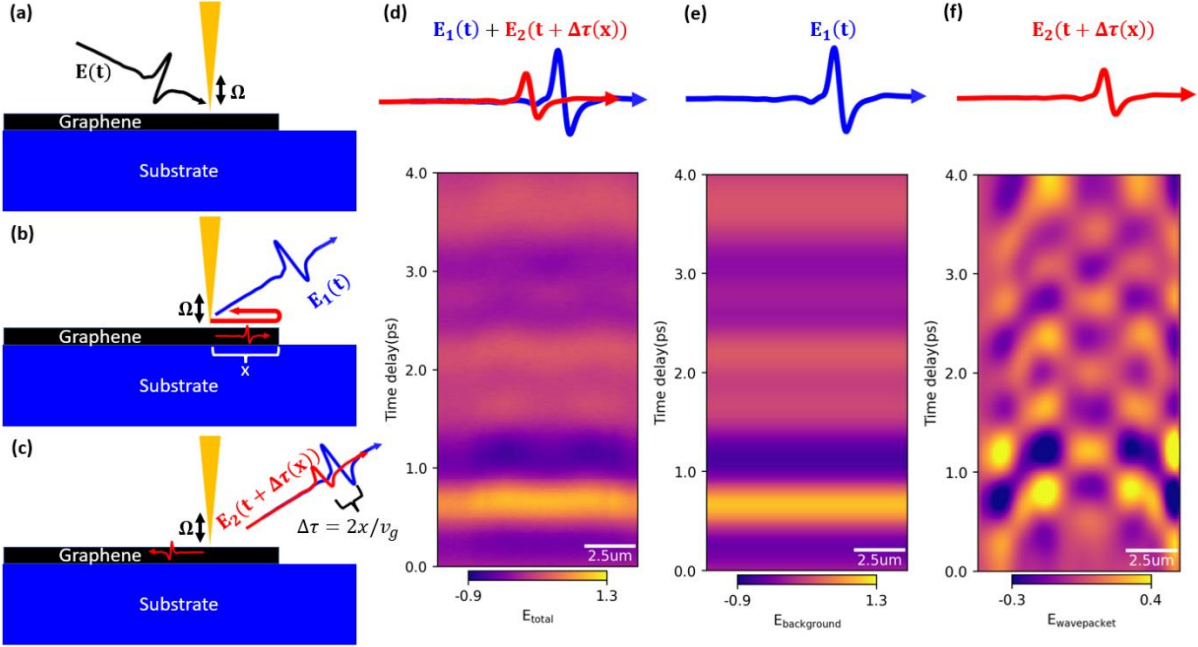

Fig. S4. Sequential representation of the detection process of nano-THz spacetime maps. (a) A THz pulse is focused onto the tip and sample. (b) $E_1(t)$ is directly scattered by the tip; meanwhile, an SPP wave packet (red) is launched by the tip and propagates in graphene. (c) Upon completing a round trip between the tip and the sample edge, the SPP wave packet is scattered by the tip, producing a trace $E_2(t)$ that co-propagates with $E_1(t)$ in free space after a time delay. The time delay ($\Delta\tau$) is determined by both the SPP group velocity and the relative distance between the tip and the edge x. (d) Schematics of the signal composition of the raw data of a spacetime map and the corresponding experimental data. (e) Schematics of the background signal, which can be estimated as the time-dependent spatially averaged THz signal inside graphene. (f) The spacetime map $E_2$, which records the worldline of SPP wave packets in the cavity.

In this section, we discuss step by step the experimental collection process of nano-THz spacetime maps. In Fig. S4(a), we display an incoming THz pulse incident on the tip-sample system. The tip oscillates at frequency $\Omega$ in the vertical direction. The tip-scattered signal contains two contributions: $E_1(t)$ and $E_2(t)$. The pulse $E_1(t)$ is directly scattered by the tip and is therefore position independent. $E_1(t)$ is representative of the local response of graphene and may be modeled by assuming an infinitely large graphene sample. The second contribution, $E_2(t)$, is non-local. The $E_2(t)$ signal is the result of the SPP wave packet being launched by the tip and completing the roundtrip between the tip and the sample edge as illustrated in Fig. S4(b). The propagating SPP wave packet is outcoupled by the tip and sent to the far field. Both $E_1(t)$ and $E_2(t)$ are demodulated at the tip tapping frequency and co-propagate in free space. The signal $E_2(t)$ is delayed with respect to $E_1(t)$ by the time that the SPP wave packet takes to travel between the tip and the sample edge. The time delay is determined by the SPP group velocity ($v_g$) and the distance between the tip and the sample, which can be expressed as $\Delta\tau = 2x/v_g$.

In Fig. S4(d), we presented the raw data for a representative spacetime map: the raw data present a superposition of the local response $E_1(t)$ and the spatially dependent polaritonic response $E_2(t + \Delta\tau(x))$. Experimentally, t is controlled by the time delay between the femtosecond probe pulse and the scattered THz pulse, which defines the vertical axis in spacetime maps. The background signal can be approximated as the spatially averaged signal inside the graphene. By taking second spatial derivatives (which will be discussed in detail in the next section), the local response is removed and the contrast is dominated by the polaritonic response (Fig. S4(f)).

In the above discussion, we focused exclusively on the tip-launched polariton. While the graphene edge may also initiate polariton wave packets, their launching efficiency of the edge at THz frequencies is diminished, leading to predominantly tip-launched modes in our observations. We posit that with an optimized launcher geometry and illumination setup, edge or contact launching modes could be viable. Typically, tip-launched and edge-launched coexist, resulting in spacetime maps featuring worldlines with slopes of $1/v_g$ and $2/v_g$.

## S6 Second derivatives for visualizing the plasmon wave packet

In this section, we discuss the background suppression/elimination method suitable for processing spacetime maps. In Section S5, we show that the measured near-field signal can be decomposed as a spatially dependent component and an independent part. The near-field signal of a spacetime map for a plasmonic medium can be expressed as

$$NF(x,t) = \sum_\omega A(\omega)e^{i(\omega t - \frac{\tilde{q}(\omega)x}{2})} + \sum_\omega B(\omega)e^{i\omega t} \qquad (S1)$$

Here, we introduce a reference-free method for removing the spatially independent component. By calculating the second spatial derivative of $NF(x,t)$, we obtain:

$$-\frac{\partial^2 NF(x,t)}{\partial x^2} = \sum_\omega \tilde{q}(\omega)^2 A(\omega)e^{i(\omega t - \frac{\tilde{q}(\omega)x}{2})} \qquad (S2)$$

In contrast to the origin signal (S1), the second derivative removes the spatially independent part. The prefactor, denoted $(A(\omega))$ in the first term in (S1), is modified to become $\tilde{q}(\omega)^2 A(\omega)$, implying a transformation of the shape of the propagating wave packet. However, the group velocity and lifetime can be unambiguously extracted from the second spatial derivatives of the spacetime map. The second spatial derivative of the raw data is shown in Fig. S5(b). After effectively eliminating the background signal, we observe a distinct checkerboard pattern, which vividly illustrates the multireflection of polariton wave packets in a plasmonic cavity.

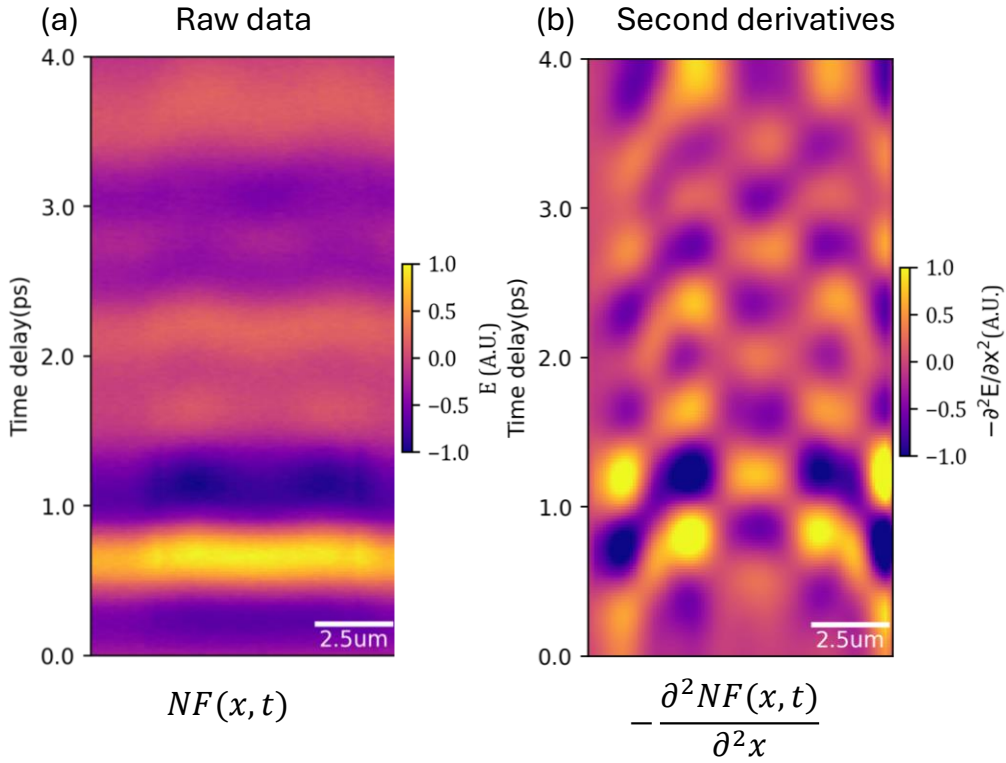

Fig. S5 Wave packet information extraction. (a) Raw data for a spacetime map, taken at $T = 55K$, $V_{bg} = 15V$. (b) Second spatial derivative of the raw data in (a).

## S7 Numerical simulations of the spacetime maps

In this section, we discuss the numerical simulations of the spacetime maps. The real-space simulations of surface plasmon polaritons is based on a computational framework which was applied to simulate plasmon response in WTe$_2$(*17*, *18*) and generate training data for polaritonic convolutional neural network(*55*). In this computational framework, the near-field response of a plasmonic medium is simulated, assuming continuous wave excitation.

In this work, we have extended the framework to accommodate pulsed excitation, taking into account broadband THz pulses that stimulate a plasmonic medium characterized by an energy-momentum dispersion ω(q). This approach allows direct comparison between experimental observations and numerical simulations. In Fig. S6, we summarize the simulation process of a spacetime map. In Fig. S6(a) and (b), we present a simulated near-field scattering signal for a square graphene sheet. The near-field signal is approximated as the z-polarization of a dipole that is raster scanned over the sample surface. A detailed description of the simulation method can be found in our previous works(55). The properties of plasmon polaritons at frequency $\omega_0$ in the simulation are determined by the wavelength $\lambda$ and the quality factor Q. The simulation results presented in Fig. S6(a) and (b) are based on continuous wave excitation. As such, they do not incorporate temporal features.

In Fig. S6(c) and (d), we present a single-cycle THz pulse in the time domain and in the frequency domain, respectively. The $E(t)$ and $E(\omega)$ traces are connected by the Fourier transformation:

$$E(t) = \frac{1}{\sqrt{N}} \sum_n \tilde{E}(\omega_n) e^{i\omega_n t} \tag{S3}$$

$$\tilde{E}(\omega) = \frac{1}{\sqrt{N}} \sum_n E(t_n) e^{-i\omega t_n} \tag{S4}$$

where $E(t)$ is a real function that describes the electric field profile in the time domain, and $\tilde{E}(\omega)$ is a complex function that determines the amplitude and phase of the pulse at different frequency components. A THz pulse in the time domain can be regarded as the superposition of continuous waves at various frequencies with different amplitudes and phases. One continuous wave component at frequency $\omega_n$ corresponds to a near-field signal distribution in the simulation volume $NF(x, y, \omega_n)$. The near-field signal for the sample in the time domain $NF(x, y, t_i)$ produced by an excitation pulse profile $\tilde{E}(\omega)$ can be calculated as the Fourier transform of $\widetilde{NF}(x, y, \omega_n) \times \tilde{E}_{\omega_n}$:

$$NF(x, y, t_i) = \sum_n \widetilde{NF}(x, y, \omega_n) \times \tilde{E}(\omega_n) \times e^{i\omega_n t_i} \tag{S5}$$

In Fig. S6(e), we present a simulated hyperspectral line scan of the near-field signal from the center to the edge along the dashed line in Fig. S6, panels (a) and (b). The dispersion of the plasmon polaritons is assumed to be linear and the group velocity is assumed to be 5.6 µm/ps. In Fig. S6(f), we show the near-field signal in the time domain based on the formula above. The worldline of a surface plasmon polariton can be visualized in the second spatial derivatives of $NF(x, y_0, t)$.

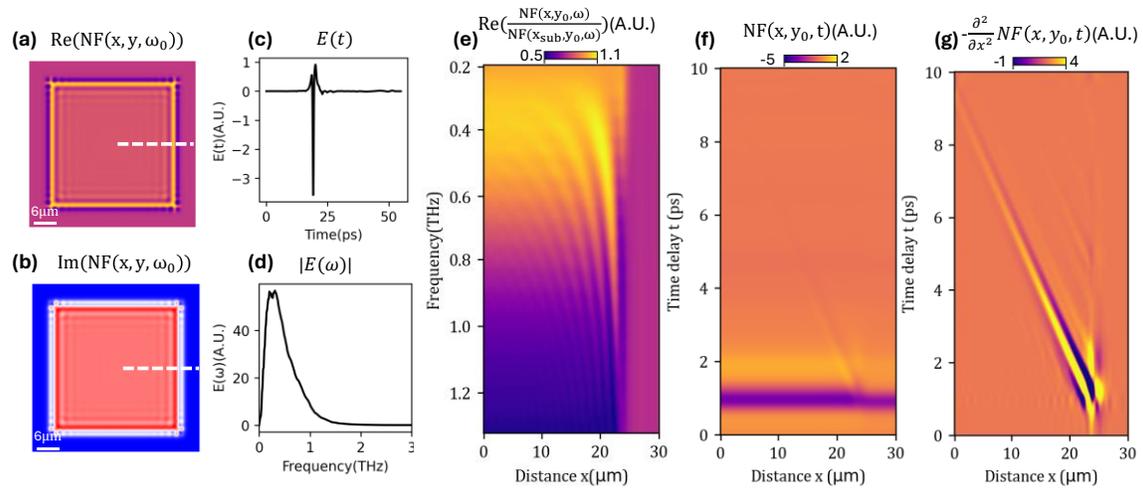

Fig. S6. Numerical simulations of spacetime maps for surface plasmon polaritons. (a) Real part and (b) imaginary part of a simulated near-field profile for a square graphene sheet under continuous wave excitation. (c) Time domain trace and (d) frequency domain spectrum of the incident THz pulse. (e) Simulated hyperspectral line scan from the center of the graphene sheet to the edge, indicated by the white dashed line in (a) and (b). The hyperspectral line scan is normalized by the near-field signal produced by the substrate. (f) Simulated spacetime map collected along the white dashed line marked in (a) and (b). (g) Spatial second derivatives of the spacetime map.

To emulate the geometry of the graphene ribbon in the experiment, we focus our numerical simulations on the ribbon geometry shown in Fig. S7.

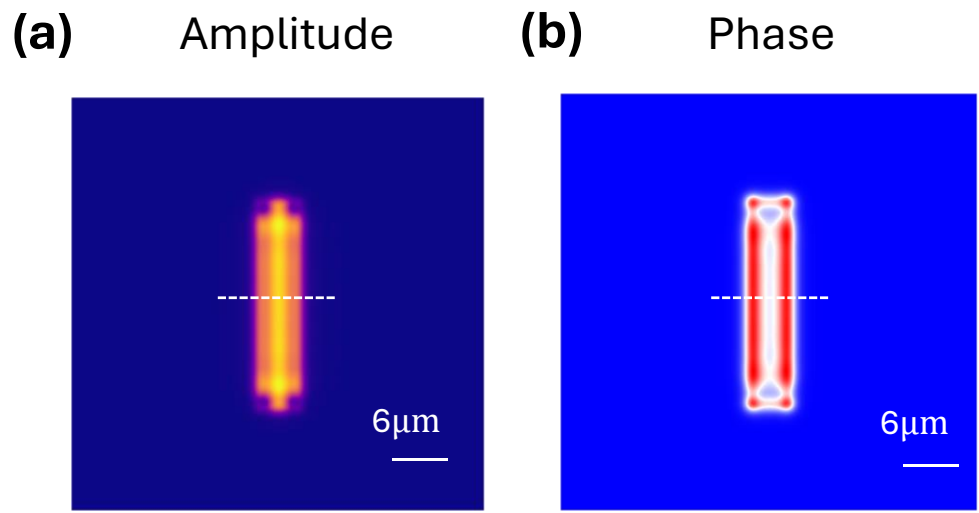

Fig. S7 Simulated near-field signal on the graphene micro-ribbon. The simulated spacetime maps for the graphene micro-ribbon are performed along the white dashed line.

# S8 Numerical simulations of the gate voltage dependence of nano-THz amplitude and phase profiles.

To understand the gate voltage dependence of the nano-THz amplitude and phase profiles, we numerically simulated the gate-dependent signals based on the plasmonic simulator introduced in Section S7. The width of the ribbon in the simulation is 6µm and the polariton wavelength and quality factor at each back gate voltage are calculated based on the dielectric function and 2-component conductivity model described in Section S12. The back gate voltage-dependent wavelength and quality factor are plotted in Fig. S8(e) and (f), respectively. Our numerical simulations adequately capture the gross features of the data in Fig.1 in the main text, which are also shown in Fig. S8.

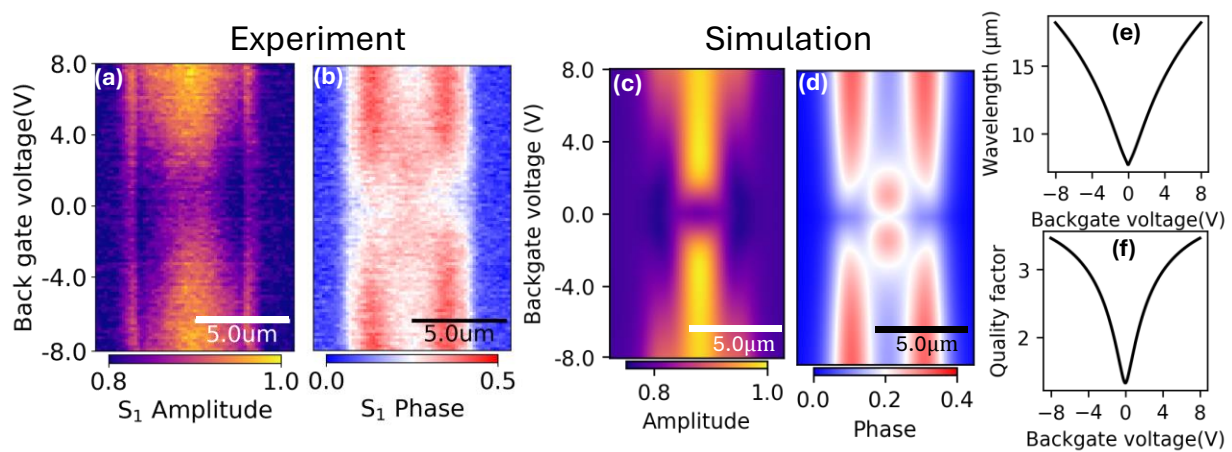

Fig. S8 The gate voltage dependence of nano-THz amplitude and phase profile across a graphene ribbon. (a-d) Experimental data are presented in Panels a,b and simulations are displayed in panels c,d. (e,f) back gate voltage-dependent wavelength and quality factor used in the numerical simulation.

## S9 Numerical simulation of the lifetime-dependent spacetime map

In this section, we describe the numerical simulations of spacetime maps with different polaritonic lifetimes:

$$\tau_p = Q_p/\omega \tag{S7}$$

Because of the complexity of the tip antenna resonance in the THz frequency range, the oscillation of the plasmon in the time domain cannot simply be fit by a sinusoidal oscillation with exponential decay. To quantitatively extract the lifetime of the SPP, we numerically simulated the SPP spacetime map at the same carrier density with different lifetimes while considering the tip resonance. In both the simulation and the experiments, the geometrical loss of the SPP is involved. Since the tip launches a circular wave, the electric field strength naturally decays with distance because of the diffusion of electromagnetic energy. In Fig. S9(a), we show a THz pulse in the time domain, which is used as the optical excitation in the numerical simulation. The time-domain trace is collected on the gold contact of the sample. The simulation is performed on a graphene ribbon whose width is $6\mu m$, and the spacetime map is collected across the ribbon (indicated as the white dashed line in Fig. S9(b)). In Fig. S9(c), we showcase a comparative analysis between the experimental spacetime maps and their corresponding simulations, which are optimized to match the lifetimes as closely as possible. A high level of agreement between experiment and simulation suggests that the numerical simulation effectively captures the nano-THz electrodynamics of the SPP in the cavity. In Fig. S9(d), we present the simulated temporal profile of the SPP oscillation extracted at the center of the cavity (white dashed line in Fig. S9(c)) from short (red) to long (blue) lifetimes. The black curve at the bottom is the time domain trace that launches the SPP in the cavity. When the SPP lifetime is short, the temporal profiles at the center of the cavity exhibit a high degree of similarity to the excitation pulse. As the SPP lifetime increases, the similarity between the temporal fields at the center of the cavity and the excitation pulse diminishes. This is because the earlier-launched SPPs persist within the cavity for a longer duration, thereby altering the temporal profiles of the signal at the cavity's center over time.

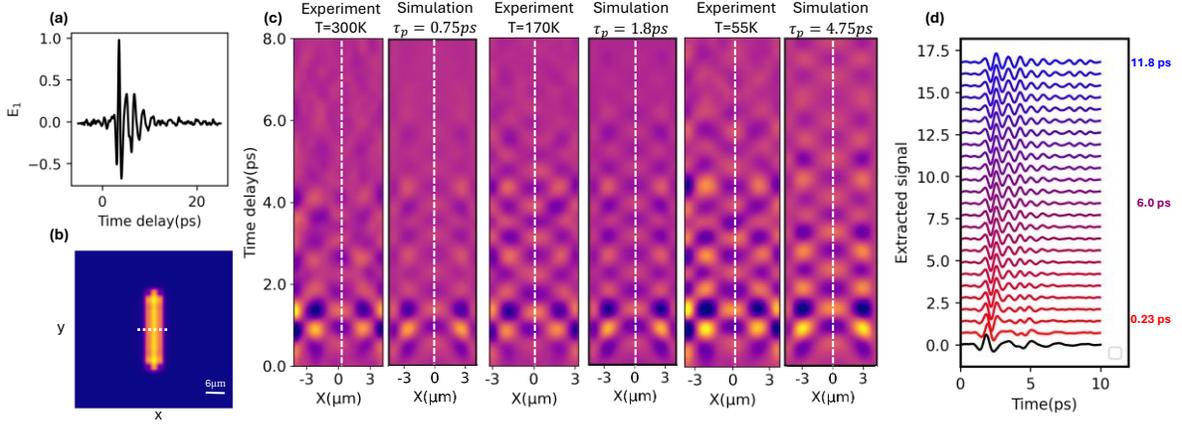

Fig. S9. Numerical simulations of the lifetime dependent spacetime maps. (a) Near-field terahertz time domain trace collected on gold contact. The near-field signal is demodulated at the 1st harmonics of the tip tapping frequency. This time domain trace serves as the excitation that initiates the SPP. (b) Geometry of the graphene ribbon in the numerical simulation. (c) Comparison between the experimental data and numerical simulations at various temperatures. (d) Simulated lifetime-dependent time domain cavity resonance. The time domain signals are extracted from the center of the cavity in the second spatial derivatives of the spacetime map (marked by the white dashed line in c). The black solid curve at the bottom is the time domain trace of the excitation pulse in the simulation.

In Fig. S10, we present a comprehensive comparison of the temporal cavity oscillations in the experimental observations and in the numerical simulations. The solid curves in Fig. S10(a-c) represent the temporal oscillations of the SPP signal, which were extracted from the center of the cavity (white dashed lines in Fig. S9(c)). The black dashed curves illustrate the numerical simulated temporal cavity oscillations with SPP lifetimes that best match the experimental data. Based on the great agreement between the experimental observations and simulated results, we conclude that the SPP lifetime is 0.75ps, 1.8ps and 4.75ps at 300K, 170K and 55K, respectively.

To quantitatively compare the experimental and simulated spacetime maps, we extracted the oscillation profiles of the near-field signal at the cavity centers for both experimental data $S_{exp}(t)$ and simulated data $S_{sim,t_P}(t)$. We define the absolute error between the experimental data and the simulated data as:

$$\Sigma_{t_P} = \int_0^8 \left| S_{exp}(t) - S_{sim,t_P}(t) \right| dt$$

The calculated absolute errors, as functions of the simulation lifetime, are depicted in Fig. S10(d-f). For each set of experimental data, the extracted lifetime is determined by the position of the local minimum. The associated errors are quantified by the variations between these minima.

The solid curves in Fig. S10(a-c) represent the temporal oscillations of the SPP signal, which were extracted from the center of the cavity (white dashed lines in Fig. S9(c)). The black dashed curves illustrate the numerical simulated temporal cavity oscillations with SPP lifetimes that best match the experimental data. Based on the great agreement between the experimental observations and simulated results, we conclude that the SPP lifetime is $0.72 \pm 0.08$ ps, $1.8 \pm 0.15$ ps and $4.75 \pm 0.35$ ps at 300K, 170K and 55K, respectively.

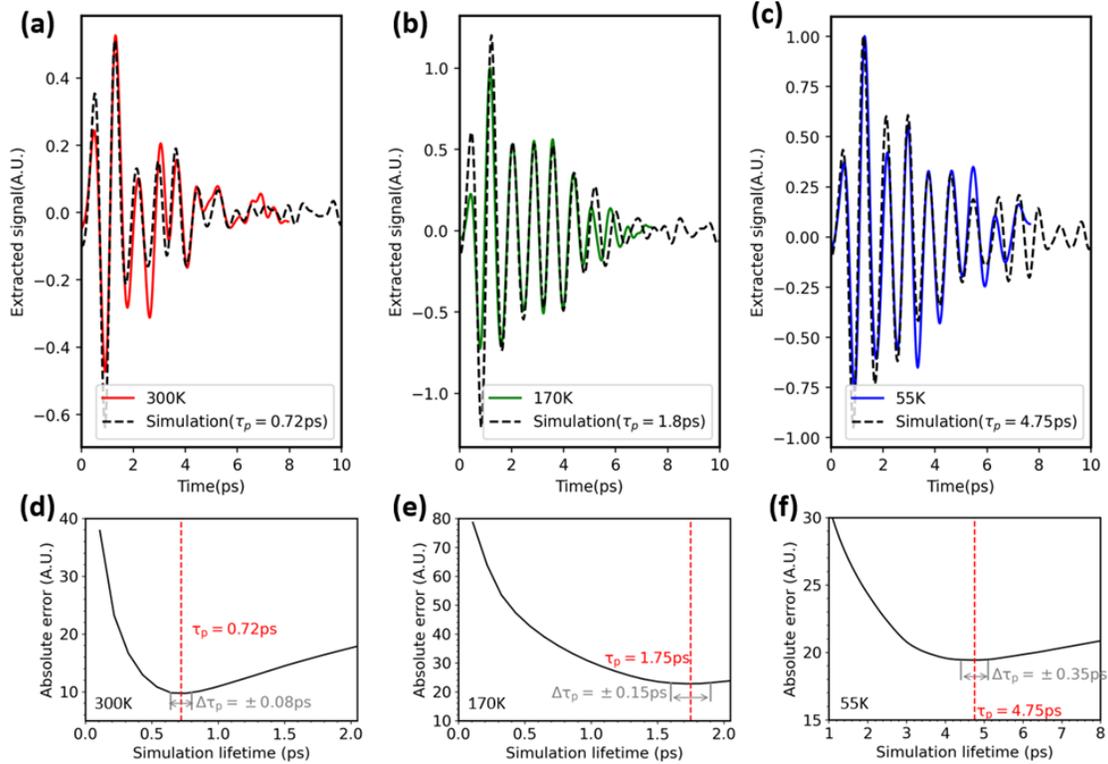

Fig. S10. A comprehensive comparison of the temporal oscillations within the cavity obtained from both experimental data and numerical simulations. The solid curves in (a-c) represent the temporal oscillations of the SPP signal extracted from the center of the cavity (as indicated by the white dashed line in Fig. S9(c)) at 300K(a), 170K(b) and 55K(c). The black dashed lines depict the simulated temporal oscillations of the SPP signal, using an SPP lifetime of 0.75ps(a), 1.80ps(b) and 4.75ps(c). The curves in (d-f) illustrate the absolute values discrepancies between experimental and simulated temporal oscillations of SPP signals as functions of simulation lifetime at 300K(d), 170K(e) and 55K(f). The lifetimes are determined by the positions of local minima, while the associated errors are quantified by the minima's widths.

# S10 Drude conductivity and acoustic plasmon dispersion

In this section, we describe the Drude model and the analytical dispersion calculations for surface plasmon polaritons.

The complex conductivity $\sigma(\omega)$ due to intraband transitions in graphene is adequately described by the Drude model(26):

$$\sigma_{intra}(\omega) = \frac{i}{\pi}\frac{D}{\omega + i\Gamma_d} \tag{S8}$$

where the Drude weight is given by:

$$D = \frac{2e^2}{\hbar^2}Tln(2\cosh(\frac{\mu}{2T})) \tag{S9}$$

We calculate the dispersion relationship of the surface plasmon polariton using local capacitance approximation(25). We obtain the dispersion by solving the pole of the 2D dielectric function:

$$\epsilon = 1 + \frac{iq^2V\sigma}{e^2\omega} \tag{S10}$$

where the Coulomb interaction is replaced by $V = e^2/C$. The capacitance is defined as:

$$C = \frac{\epsilon}{4\pi d} \tag{S11}$$

In all calculations, we used $\epsilon = 5$ and d=300nm.

The dispersion of the SPP can be calculated by solving the pole of the 2D dielectric function:

$$1 + \frac{iq^2V\sigma}{e^2\omega} = 0 \tag{S12}$$

The dispersion of the SPP can be expressed as the frequency-dependent complex wave vector:

$$q_1 + iq_2 = \omega\sqrt{\frac{\pi C}{D}}\sqrt{1 + i\frac{\Gamma_d}{\omega}} \tag{S13}$$

The above formula is the complete form of the SPP dispersion and is used for all the numerical calculations in the manuscript.

When $\Gamma_d/\omega \ll 1$, we find

$$q_1 \approx \omega\sqrt{\frac{\pi C}{D}} \tag{S14}$$

$$q_2 \approx \frac{1}{2}\sqrt{\frac{\pi C}{D}}\Gamma_d \tag{S15}$$

Based on the definition of the group velocity, quality factor and lifetime:

$$v_g = \frac{d\omega}{dq} \tag{S16}$$

$$Q = \frac{q_1}{q_2} = \frac{2\omega}{\Gamma_d} \tag{S17}$$

$$\tau_p = \frac{Q}{\omega} \tag{S18}$$

one can obtain the relationship between the geometrical metrics of SPP worldlines and the Drude model parameters, as summarized in Eq.2 in the main text.

# S11 Temperature-dependent damping of SPPs in the Fermi liquid regime

In this section, we describe various scattering mechanisms that are relevant to the dissipation of SPP in the Fermi liquid region ($\mu \gg k_b T$).

Electron-phonon scattering rate

The electronic scattering rate in graphene due to scattering on acoustic phonons is given by(*29*):

$$\Gamma_{ph} = \frac{1}{\hbar^3} \frac{\beta_A^2 |\mu|}{\mu_s v_F^2} \left( \frac{1}{v_l^2} + \frac{1}{v_t^2} \right) k_b T \tag{S19}$$

where $\beta_A = 5.0 eV$ is the pseudo-magnetic field coupling constant, $v_l$ and $v_t$ are the velocities of longitudinal and transverse acoustic phonons in graphene, $v_f$ is the Fermi velocity and $\mu_s$ is the graphene mass density. The parameters used in the numerical calculation are recorded in Table S1.

Table S1. Numerical parameters for calculating the electron-phonon scattering rate in graphene(*29*)

| | | |
|---|---|---|
| $\beta_A$ | Pseudo-magnetic field coupling constant | $5.0 eV$ |
| $v_l$ | Velocity of longitudinal acoustic phonons | $2.2 \times 10^6 cm\ s^{-1}$ |
| $v_t$ | Velocity of transverse acoustic phonons | $1.4 \times 10^6\ cm\ s^{-1}$ |
| $v_f$ | Fermi velocity | $1.3 \times 10^8 cm\ s^{-1}$ |
| $\mu_s$ | Graphene mass density | $7.6 \times 10^{-8} g\ cm^{-2}$ |

Electron-impurity scattering rate

The electron-impurity scattering rate in graphene is given by(*56*):

$$\Gamma_{imp} = \frac{1}{\hbar} \frac{\left(\frac{Ze^2}{\epsilon}\right)^2 \rho_{imp}}{\max[k_B T, \mu]} \tag{S20}$$

The electron-impurity scattering rate is proportional to the density of impurities in graphene. For the high-quality hBN-encapsulated graphene sample, we use $\rho_{imp} = 2.1 \times 10^9 cm^{-1}$, which has been reported in previous on-the-chip terahertz spectroscopy measurements(8). The contribution of SPP damping from electron-impurity scattering is negligible in the parameter space pertinent to our experiments.

## S12 Extraction and uncertainties of group velocity

In this section, we discuss the extraction procedure of SPP group velocity in a spacetime map collected on a semi-infinite SPP medium or an SPP cavity. Here, we consider that the SPP wave packet is launched and detected by the tip.

In Fig. S11(a), we show one simulated spacetime map on a semi-infinite sample. The group velocity ($V_g$) can be determined from the gradient of the worldline: specifically, $V_g$ equals twice the change in position ($\Delta x$) divided by the change in time ($\Delta t$).

In Fig. S11(b), we present a simulated spacetime map on a cavity with 6-$\mu m$ width, which holds SPPs with a group velocity identical to that depicted in Fig. S11(a). The group velocity can be extracted from the slope of the worldline, marked by the white circle in Fig. S11(b). Because the signal contrast in the spacetime map varies between the edge and the center of the cavity, the best location for extracting the slope of the worldline is at the location marked by the blue dashed line in Fig. S11(c). We zoom in on the region marked by the white dashed rectangle in Fig. S11(b). The black dashed line indicates the position of the maximum value in the spatial direction. A segment of the dashed line, situated at the one-quarter point within the cavity, exhibits a constant slope and is consistent with the slope determined in Fig. S11(a).

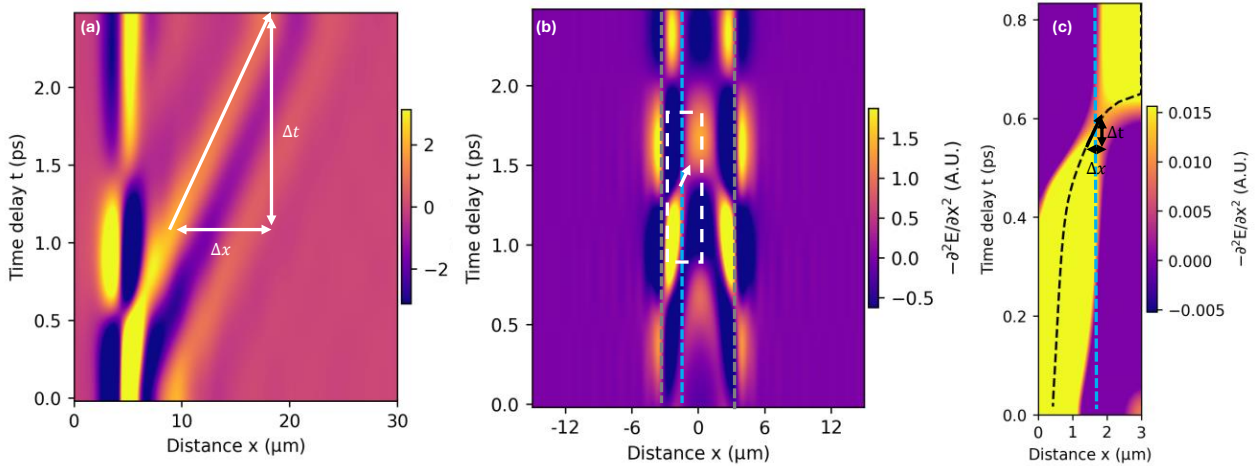

Fig S11. (a) Simulated spacetime map of a semi-infinite sample. The slope is well defined by the SPP detection worldline. (b) Simulated spacetime map of a plasmonic cavity with 6-μm width. The gray dashed lines indicate the boundary of the cavity. The group velocity can be effectively extracted from the location marked by the blue dashed line. (c) Zoomed-in view of the region in (b) marked by the white dashed rectangle. The black dashed line shows the trace of the maximum value along the space axis. At the one-

quarter point of the cavity, the slope of the dashed line is constant and is the same as the one extracted from the SPP detection worldline in the semi-infinite sample (a).

Here, we apply the same strategy to extract the group velocity and define its error on experimental data. In Fig. S12(a), we present the second derivative of a spacetime map, which was collected across the graphene ribbon at room temperature, with a back gate voltage set at 4V. In Fig. S12(b), we zoom in on the region in Fig. S12(a) marked with the white dashed rectangle. The white dots indicate the locations of the maximum values in the spatial direction at different time delays. The group velocity of the SPPs is represented by a straight-line segment with a constant slope, which is clearly illustrated in Fig. S12(c). Based on the definition of group velocity of a tip-launched SPP, we extract the group velocity using the formula $v_g = \frac{2\Delta x}{\Delta t}$. The error of the extracted group velocity is determined by the spatial ($\sigma_x$) and temporal($\sigma_t$) resolution of the spacetime map: $\sigma_{v_g} = \frac{\sigma_x}{\Delta t} + \frac{\Delta x}{\Delta t^2}\sigma_t$. The gate-dependent group velocity presented in Fig. 4(d) in the main text is the average of the slopes extracted from the left and right sides of the graphene ribbon.

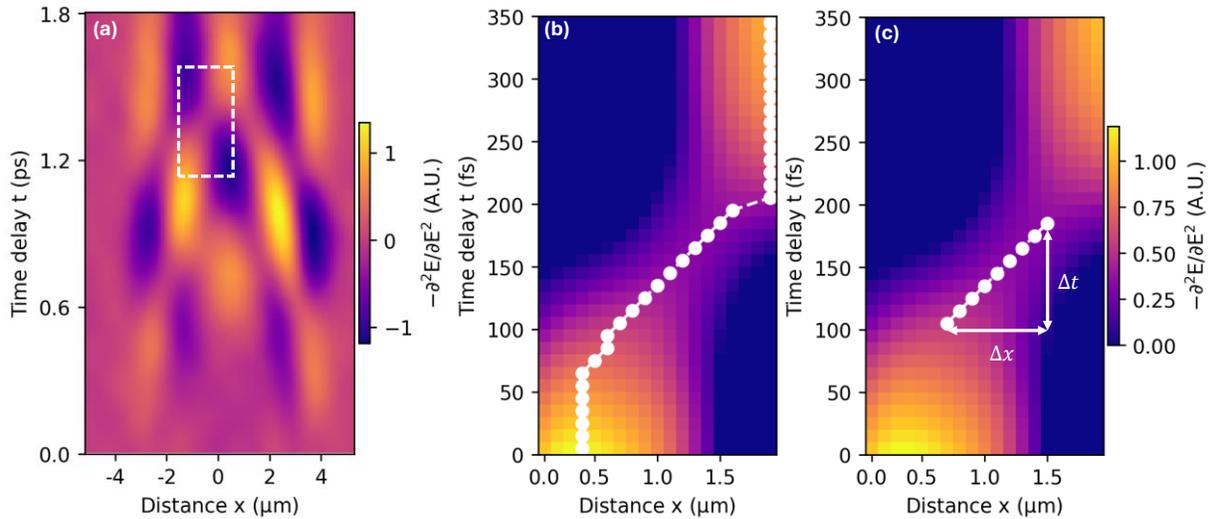

Fig. S12 Group velocity extraction and error estimation on experimental data. (a) Experimental data plotted in the form of a spacetime map measured at $V_{bg} = 4V$. (b) Zoomed-in view of the region in (a) marked with the white dashed rectangle. The white dots show the trace of the maximized signal extracted at each time delay. A straight-line segment at the one-quarter point of the cavity is observed. (c) Visualization of the slope of the SPP worldline in an SPP cavity. The error of the group velocity is determined by the spatial and temporal resolution on a spacetime map.

# S13 Interband transitions

In this section, we discuss the interband contribution to the conductivity of graphene. As the doping is brought closer to charge neutrality, interband contributions could potentially become important even at THz frequencies. However, we have found that at room temperature, the contribution of interband processes is negligible compared to that of the intraband (Drude) response.

We evaluate the interband conductivity following standard methods in the literature(*57–59*). An analytical expression for the optical conductivity in the local approximation reads as:

$$\sigma_{inter}(\omega) = \frac{e^2}{h}\frac{\pi}{2}\left(g(\omega) + \frac{4i\omega}{\pi}\int_0^\infty \frac{g(2\omega') - g(\omega)}{\omega^2 - 4\omega'^2}d\omega'\right) \quad (S21)$$

where the function g is defined as:

$$g(\omega) = \frac{\sinh\left(\frac{\omega}{2T}\right)}{\cosh\left(\frac{\mu}{T}\right) + \cosh\left(\frac{\omega}{2T}\right)} \quad (S22)$$

Here $\omega$, $\mu$, and $T$ are frequency, chemical potential, and temperature, respectively.

In the T→0 limit, the real part of this expression becomes $\left(\frac{e^2}{h}\frac{\pi}{2}\right)\Theta(\omega - 2\mu)$, i.e., the well-known universal optical conductance of monolayer graphene(*60*).

In Fig. S13, we show the temperature-dependent interband contribution to the graphene optical conductivity based on (S21) and (S22) in the frequency range relevant to our experiments. The magnitude of $\sigma_{inter}$ increases as temperature decreases. In Fig. S13(b-e), we show the intraband and interband contributions at the charge neutral point. At room temperature, intraband transitions dominate the conductivity. The interband contribution begins to become comparable to the intraband contribution below 30K, at which point the transition energy (4mev) is comparable to the thermal energy ($k_b T$).

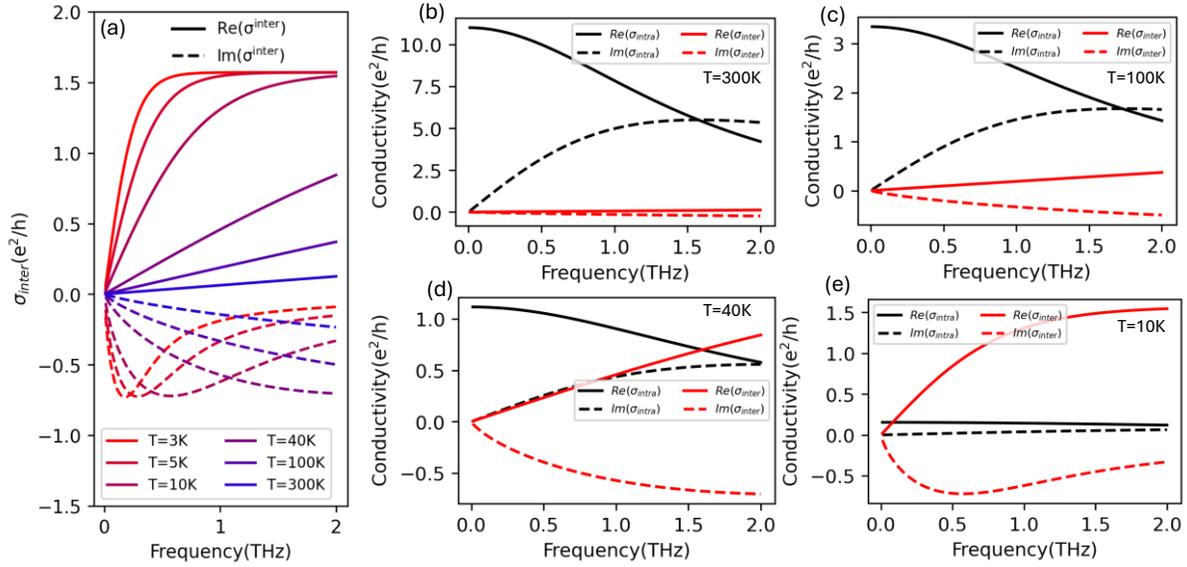

Fig. S13 Interband transition contribution to the graphene optical conductivity. (a) Interband optical conductivity of an undoped graphene sample calculated at different temperatures. (b-e) Comparison between intraband and interband optical response at 300K, 100K, 40K and 10K.

## S14 Two-component conductivity model

The two-component conductivity model was reported in Ref.(*36*):

$$\sigma(\omega) = \frac{i}{\pi}\frac{D_h}{\omega + i\Gamma_d} + \frac{i}{\pi}\frac{D - D_h}{\omega + i\Gamma_d + i\Gamma_e} \tag{S23}$$

where $D_h$ is the hydrodynamic Drude weight and D is the kinetic Drude weight.

The kinetic Drude weight is given by

$$D = \frac{2e^2}{\hbar^2} T \ln(2\cosh(\frac{\mu}{2T})) \tag{S24}$$

The hydrodynamic model is given by

$$D_h = \pi e^2 n / m_h \tag{S25}$$

where the corresponding hydrodynamic mass is given by

$$m_h = \frac{3}{\pi}\frac{T^3}{\hbar^2 v_F^4 n}\left[\frac{\pi^2}{3}\frac{\mu}{T} + \frac{1}{3}\frac{\mu^3}{T^3} - 4Li_3(-e^{\mu T})\right] \tag{S26}$$

$Li_n(z)$ is the polylogarithm function of order n and argument z.

At the charge neutrality point, $D_h = 0$ and the graphene is in the Dirac fluid regime is marked by a notable contribution from the electron-hole scattering:

$$\sigma(\omega) = \frac{i}{\pi}\frac{D_k}{\omega + i\Gamma_d + i\Gamma_e} \quad (\mu = 0 mev) \tag{S27}$$

At high carrier density, $D_h$ is approximately the same as $D_k$ and the graphene conductivity collapses into the non-interacting model in which the electron-hole scattering is dormant in the electromagnetic response:

$$\sigma(\omega) = \frac{i}{\pi}\frac{D_k}{\omega + i\Gamma_d} \quad (\mu \gg k_b T) \tag{S28}$$

# S15 Geometrical decay of Surface plasmon Polariton field strength

In Fig. 4 in the main text, we presented three representative spacetime maps collected at $V_{bg}$=0, 1, and 5V along the long axis of the graphene ribbon. A longer distance in space allows us to track the decay of the free propagating wave packet without considering intersection and multireflection. Since the wavelength of the SPP exceeds the width of the ribbon, the geometrical decay of SPP field strength resulting from its spreading across space can be disregarded. The temporal decay of the SPP wave packet can be described by an exponential decay function $e^{-t/\tau_p}$, where $\tau_p$ is the SPP lifetime.

In the following numerical simulations, we demonstrate that the geometrical decay is absent in the experimental scenario in Fig. 4(e-g) in the main text, where the width of the graphene ribbon is smaller than the wavelength of the SPP. In Fig. S14(a-c), we simulate the spacetime map collected on a large sample; the SPP lifetime in the simulation is set to be 1.43ps. The sample geometry is depicted in Fig. S14(a). The electromagnetic energy of an SPP wave packet initiated by a tip can propagate in both the x and y directions. In Fig. S14(c), we juxtapose the exponential decay function with the simulated field strength extracted along the SPP worldline. The worldline is denoted by the black dashed line in Fig. S14(b).

The simulated field strength on the SPP worldline exhibits a decay rate that surpasses that of the exponential decay function. This accelerated decay can be attributed to additional geometric factors that need to be considered.(*61*)

In the data acquisition scheme for Fig. 4(e-f) in the main text, the width of the ribbon is indeed smaller than the SPP wavelength (defined as 1THz). This implies that the sample geometry employed in the simulations for Fig. S14(a-c) may not be applicable. In Fig. S14(d-f), we numerically simulate the spacetime map in a graphene ribbon. The sample geometry is shown in Fig. S14(d); the width of the graphene ribbon is 6μm. In Fig. S14(f), we show the temporal decay profile of the SPP field strength along the worldline on a graphene ribbon (black solid curve) and on a graphene square (red solid curve). The temporal decay along the ribbon is slower than on a square sample, and it aligns more closely with an exponential decay function. On a graphene ribbon whose width is less than the SPP wavelength, the SPP lifetime can be effectively determined through numerical fitting using an exponential decay function.

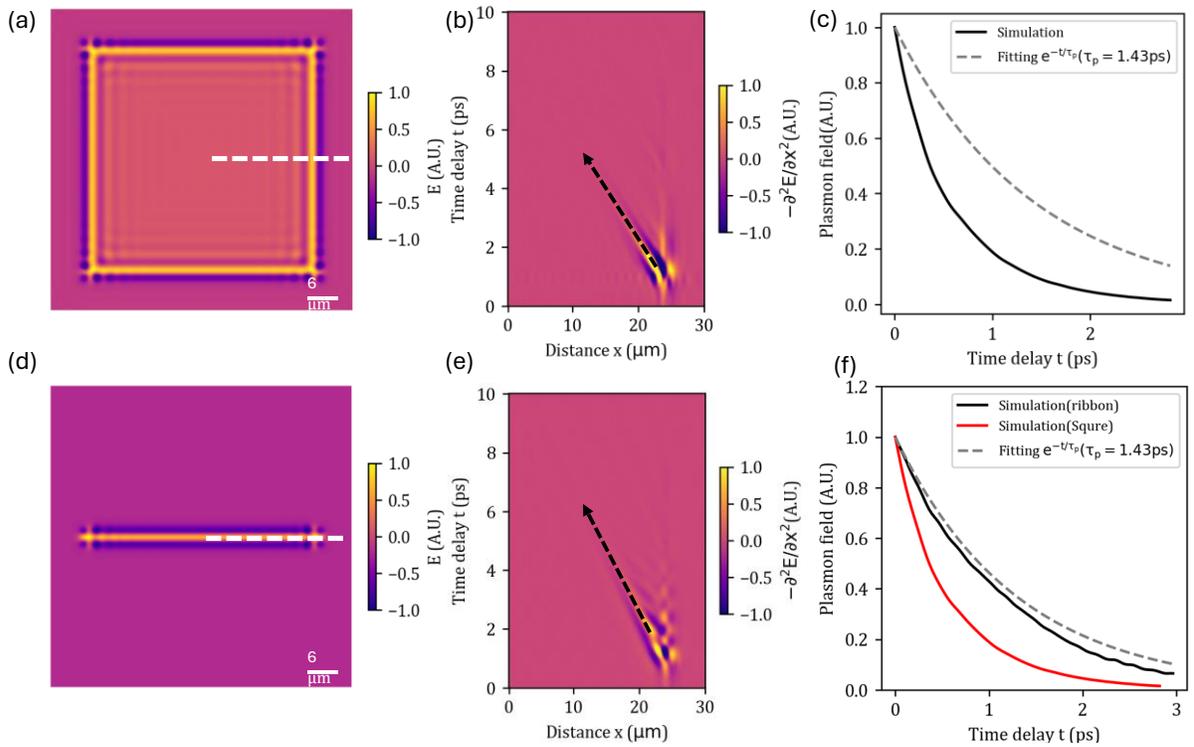

Fig. S14 Geometrical decay in SPP medium with different geometry. (a) Square medium. (b) Simulated spacetime map. (c) Extracted time-dependent electric field strength along the worldline and comparison to the fitting. (d) Graphene ribbon. (e) Simulated spacetime map along a 6μm-wide ribbon. (f) Extracted time-dependent electric field strength along the worldlines in the graphene ribbon in (d) and the graphene square in (a). The temporal decay of the SPP field strength in the graphene ribbon can be well described by the exponential decay function without considering geometrical decay.

# S16 Group velocity renormalization from a large scattering rate

The dispersion relationship for a screened SPP is given by:

$$\tilde{q} = q_1 + iq_2 = \omega \sqrt{\frac{\pi C}{D}} \sqrt{1 + i\frac{\Gamma}{\omega}} \tag{S29}$$

The approximation used to derive Equation (2) in the main text is not valid when the damping is comparable to or larger than the frequency: the presence of the damping leads to the renormalization of the SPP wave vector ($q_1$) and the group velocity. In this section, we explore the dependence of the complex SPP wave vector on the scattering rate ($\Gamma$). The real and imaginary parts of the complex wave vector $\tilde{q}$ are plotted as a function of the scattering rate in Fig. S15. The imaginary part of $\tilde{q}$ increases with $\Gamma$ monotonically. The real part of $\tilde{q}$ remains approximately constant when $\Gamma/\omega$ is small. However, it begins to deviate and increase when $\Gamma/\omega$ is comparable to or larger than 1. This explains why the real part of the wave vector increases with the scattering rate when the scattering rate is much larger than the frequency of the excitation. It follows that the group velocity decreases with the scattering rate.

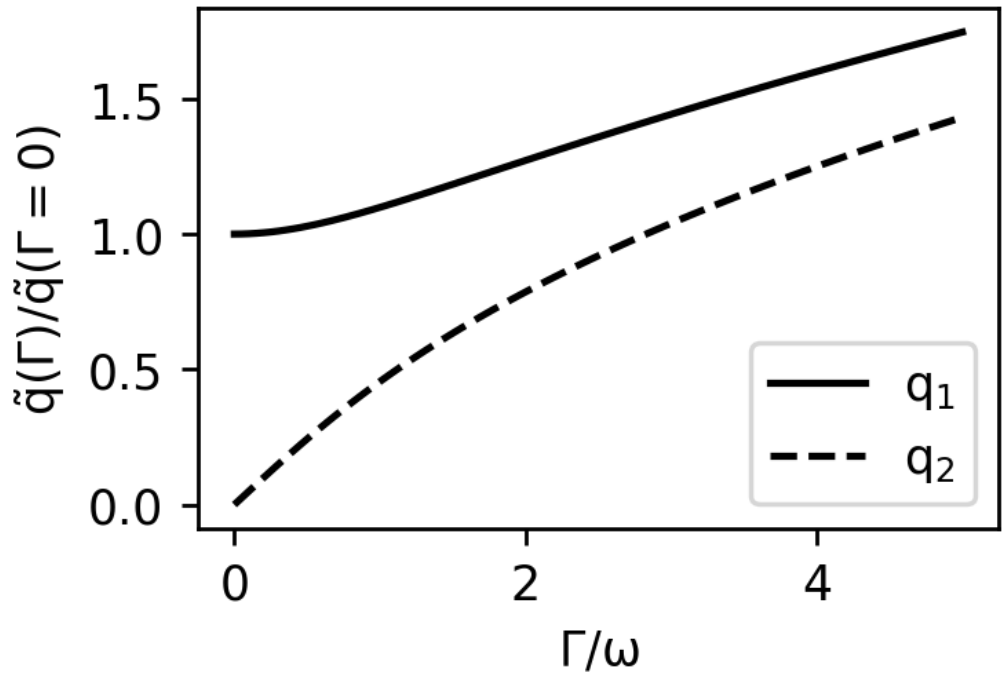

Figure S15. Dispersion renormalization in the presence of a large scattering rate. Here, we illustrate how the real part (solid line) and imaginary part (dashed line) of the SPP wave vector change with the scattering rate $\Gamma$.

## S17 Evaluating graphene's fine structure constant

We follow previous literature(*8*, *14*) to estimate graphene's fine structure constant.

The fine structure constant of graphene, denoted $\alpha$, is defined as:

$$\alpha = \frac{1}{137} \frac{c}{v_f} \frac{1}{\epsilon} \tag{S30}$$

Using the Fermi velocity $v_f = 1.3 \times 10^6 m/s$ and dielectric screening $\epsilon = 5$, we calculated the fine structure constant theoretically as $\alpha_{RPA} = 0.336$.

The extracted scattering rate at the room temperature is $\Gamma_e = 3\text{THz}$.

The quantum critical electron-electron scattering is given by:

$$\Gamma_e = C \left(\frac{k_B T}{\hbar}\right) \tag{S31}$$

The constant C extracted from the measured electronic scattering rate is C=0.43. The fine structure constant evaluated from the measured electronic scattering rate is $\alpha = \sqrt{C/3.646} = 0.343$